\documentclass[]{aa} 

\usepackage{epsfig}

\newcommand{\pl}{\partial}
 
\newcommand{\beq}{\begin{equation}} 
\newcommand{\eeq}{\end{equation}} 
\newcommand{\beqa}{\begin{eqnarray}} 
\newcommand{\eeqa}{\end{eqnarray}} 
\newcommand{\bea}{\begin{array}} 
\newcommand{\ea}{\end{array}} 
\newcommand{\lag}{\langle} 
\newcommand{\rag}{\rangle}
\newcommand{\Om}{\Omega_{\rm m0}}

\newcommand{\rhobm}{\overline{\rho}_{\rm m}}

\newcommand{\bx}{{\bf x}}

\newcommand{\dd}{{\rm d}}
\newcommand{\tdelta}{\tilde{\delta}}
\newcommand{\bk}{{\bf k}}
\newcommand{\ii}{{\rm i}}
\newcommand{\cF}{{\cal F}}
\newcommand{\cP}{{\cal P}}
\newcommand{\cD}{{\cal D}}
\newcommand{\cS}{{\cal S}}
\newcommand{\bq}{{\bf q}}

\newcommand{\tW}{\tilde{W}}

\newcommand{\fNL}{f_{\rm NL}}
\newcommand{\tPhi}{\tilde{\Phi}}
\newcommand{\tphi}{\tilde{\phi}}
\newcommand{\tchi}{\tilde{\chi}}
\newcommand{\fNLd}{\fNL^{\delta}}
\newcommand{\tfNLd}{\tilde{f}_{\rm NL}^{\delta}}

\newcommand{\Gam}{\Gamma}
\newcommand{\tf}{\tilde{f}}
\newcommand{\tfdel}{\tf^{\delta}}

\begin{document} 
 
\title{Mass function and bias of dark matter halos for non-Gaussian initial conditions} 
\author{P. Valageas}   
\institute{Institut de Physique Th\'eorique, CEA Saclay, 91191 Gif-sur-Yvette, 
France}  
\date{Received / Accepted } 
 
\abstract
{}
{We revisit the derivation of the mass function and the bias of dark matter halos
for non-Gaussian initial conditions.}
{We use a steepest-descent approach to point out that exact results can be obtained
for the high-mass tail of the halo mass function and the two-point correlation of
massive halos. Focusing on primordial non-Gaussianity of the local type, we check
that these results agree with numerical simulations.}
{The high-mass cutoff of the halo mass function takes the same form as the
one obtained from the Press-Schechter formalism, but with a linear threshold
$\delta_L$ that depends on the definition of the halo (i.e. $\delta_L\simeq 1.59$
for a nonlinear density contrast of $200$). We show that a simple formula, which
obeys this high-mass asymptotic and uses the fit obtained for Gaussian initial
conditions, matches numerical simulations while keeping the mass function normalized
to unity. Next, by deriving the real-space halo two-point correlation in the spirit
of Kaiser (1984) and taking a Fourier transform, we obtain good agreement
with simulations for the correction to the halo bias, $\Delta b_M(k,\fNL)$, due to
primordial non-Gaussianity. Therefore, neither the halo mass function nor the bias
require an ad-hoc parameter $q$ (such as $\delta_c\rightarrow \delta_c \sqrt{q}$),
provided one uses the correct linear threshold $\delta_L$ and pays attention to halo
displacements.
The nonlinear real-space expression can be useful for checking that the
``linearized'' bias is a valid approximation. Moreover, it clearly shows how the
baryon acoustic oscillation at $\sim 100 h^{-1}$Mpc is amplified by the bias of
massive halos and modified by primordial non-Gaussianity. On smaller scales, 
$30 <x< 90 h^{-1}$Mpc, the correction to the real-space bias roughly scales
as $\fNL \, b_M(\fNL=0) \, x^2$. The low-$k$ behavior of the halo bias
does not imply a divergent real-space correlation, so that one does not
need to introduce counterterms that depend on the survey size.}
{}


\maketitle

\section{Introduction} 
\label{Introduction}

Standard single-field slow-roll inflationary models predict a nearly scale-invariant
and Gaussian spectrum of primordial curvature fluctuations (e.g., Bartolo et al. 2004).
This agrees with current observations of the cosmic microwave background
(CMB) anisotropies (Komatsu et al. 2009) and of large-scale structures
(Slosar et al. 2008). Nevertheless, several inflationary models predict a
potentially observable level of non-Gaussianity (e.g., Bartolo et al. 2004 for a review),
so that constraining or detecting primordial non-Gaussianity is an important task
for current cosmological studies. This would allow one to rule out some of the many
inflationary models that have already been proposed. In particular, in many cases,
the non-Gaussianity is of the local type, meaning that it only depends on the local
value of Bardeen's potential $\Phi$. That is, the latter can be decomposed as
\beq
{\rm local \;\,  type:} \;\;\; \Phi(\bx) = \phi(\bx) + \fNL \; (\phi(\bx)^2-\lag\phi^2\rag) ,
\label{fNLdef}
\eeq
where $\phi$ is a Gaussian random field. Simple slow-roll inflation gives a
parameter $\fNL$ of $10^{-2}$, but this would be masked by the nonlinearities
that arise from the dynamics (e.g., from the nonlinearity of Einstein's equations,
see Bartolo et al. 2004) or from the physical processes involved by the observables
(e.g., perturbations at recombination that affect the CMB, see Senatore et al. 2009),
which give an effective $\fNL$ close to unity. High values of $\fNL$ can be
obtained, for instance, from multifield inflation (Bartolo et al. 2002; Lyth et al. 2003),
self-interactions (Falk et al. 1993), tachyonic preheating in hybrid inflation
(Barnaby \& Cline 2006), or ghost inflation (Arkani-Hamed et al. 2004).
Current limits are $-9<\fNL<111$ from CMB (Komatsu et al. 2009) and
$-29<\fNL<70$ from large-scale structures (Slosar et al. 2008).

The effects of primordial non-Gaussianity on large-scale structures can be seen,
for instance, through the mass function of virialized halos, especially in the
high-mass tail as the steep falloff magnifies the sensitivity to initial conditions 
(Lucchin \& Matarrese 1988; Colafrancesco et al. 1989; Grossi et al. 2007;
Maggiore \& Riotto 2009). 
This allows using the X-ray luminosity function of clusters to constrain the amount
of non-Gaussianity (Amara \& Refregier 2004).

A second probe of non-Gaussianity is provided by the clustering of these halos,
as measured through their many-body correlations. In particular, the halo two-point
correlation can be significantly increased if the underlying primordial density field 
is non-Gaussian (Grinstein \& Wise 1986).
More specifically, Dalal et al. (2008) have recently shown that primordial
non-Gaussianity of the local type (\ref{fNLdef}) gives rises to a strongly 
scale-dependent bias on large scales, whereas in the Gaussian case the bias is roughly
constant in this range. Thus, at linear order over $\fNL$ they obtain in Fourier
space a correction of the form
\beq
\Delta b_M(k,\fNL) = \fNL \; [ b_M(k,0) - 1 ]  \; 
\frac{3\delta_L\Om H_0^2}{c^2k^2T(k)D(z)} ,
\label{Dbk_Dal}
\eeq
where $b_M(k,0)$ is the Gaussian-case bias (defined as the ratio of the halo and matter
power spectra, $b_M^2(k)=P_M(k)/P(k)$, for objects of mass $M$).
Here $\delta_L$ is the linear matter density contrast associated with virialized
objects (usually taken as $\delta_L \simeq 1.686$), $T(k)$ the transfer function,
and $D(z)$ the linear growth factor, normalized as $D(z) \rightarrow (1+z)^{-1}$
at high redshift. Then, Dalal et al. (2008) checked in numerical simulations
that, in agreement with Eq.(\ref{Dbk_Dal}), the halo bias correction roughly grows as
$1/k^2$ at low $k$. This gives rise to a significant and specific signal that has
already been used to constrain $\fNL$ (Slosar et al. 2008).

In this article, following a previous work devoted to the Gaussian case
(Valageas 2009b), we revisit the derivations of the halo mass function and of the
bias for primordial non-Gaussianity. Although we focus on the local type
(\ref{fNLdef}), our approach also applies to any non-Gaussian model where
Bardeen's potential can be written as the sum of linear and quadratic terms over
an auxiliary Gaussian field, that is, where $\fNL$ becomes a convolution kernel.
(It also extends to cases that contain higher order terms and multiple Gaussian
fields.)

After introducing our notations and the quantities needed for our calculations
in section~\ref{initial-conditions}, we consider the halo mass function in 
section~\ref{Mass-function-halos}. Here, our aim is to argue that the exponential
cutoff of the high-mass tail can be obtained exactly from a saddle-point approach.
This is equivalent to the saddle-point computation of Matarrese et al. (2000),
which is often used to model the non-Gaussian halo mass function. However,
with a different treatment, we simultaneously derive the linear density profile of
this saddle point, which allows us to check that the latter is almost insensitive to
primordial non-Gaussianity, so that shell crossing is not amplified and exact results
can be obtained provided one uses the correct linear density threshold, rather than
the usual one. We also propose a simple recipe to match the dependence on $\fNL$
of the high-mass tail while keeping the mass function normalized to unity.
Then, in section~\ref{Bias-of-halos} we consider the two-point correlation of
dark matter halos in real space, following the spirit of Kaiser (1984).
Next, taking a Fourier transform we obtain the halo bias in Fourier space.
Here, our aim is to show that one does not need to introduce free parameters to
match the results of numerical simulations. Moreover, the nonlinear real-space
expression is of interest by itself and it also allows one to check whether the
``linearized'' bias is valid on the range of interest.
Finally, we conclude in section~\ref{Conclusion}.

\section{Non-Gaussian initial conditions}
\label{initial-conditions}

We focus in this paper on non-Gaussianities of the local type, where Bardeen's
potential $\Phi$ is of the form (\ref{fNLdef}), with $\phi$ a Gaussian random field.
On scales smaller than the Hubble
radius, $\Phi$ equals minus the Newtonian gravitational potential and the
Poisson equation gives in Fourier space (Slosar et al. 2008)
\beq
\tdelta_L(\bk,z) = \alpha(k,z) \tPhi(\bk) \;\;\; \mbox{with} \;\;
\alpha(k,z) = \frac{2c^2k^2T(k)D(z)}{3\Om H_0^2} ,
\label{alphadef}
\eeq
where $\delta_L$ is the linear matter density contrast,
$T(k)$ is the transfer function and $D(z)$ is the linear growth
factor, normalized as $D(z) \rightarrow (1+z)^{-1}$ at high redshift.
Unless stated otherwise, we normalize the Fourier transform as
\beq
\delta_L(\bx) = \int\dd\bk \, e^{\ii\bk.\bx} \, \tdelta_L(\bk) .
\label{Fourier_norm}
\eeq
Note that we define $\fNL$ by applying Eq.(\ref{fNLdef}) at early times
(i.e. $z=\infty$), which is sometimes called the ``CMB convention'', whereas
some authors first linearly extrapolate $\Phi$ at $z=0$ (``LSS convention'').
Thus, both conventions are related by $\fNL^{\rm CMB} = D(0) \fNL^{\rm LSS}$
(Pillepich et al. 2010). Then, defining the time-dependent Gaussian field
$\chi$ by
\beq
\tchi(\bk,z) = \alpha(k,z) \, \tphi(\bk) ,
\label{chidef}
\eeq
we can write the linear density field at redshift $z$ as
\beqa
\tdelta_L(\bk) & = &  \tchi(\bk) + \int \dd\bk_1\dd\bk_2 \, \delta_D(\bk_1+\bk_2-\bk)
\, \tfNLd(\bk_1,\bk_2) \nonumber \\
&& \times \, \tchi(\bk_1) \, \tchi(\bk_2) ,
\label{deltak_chik}
\eeqa
with
\beq
{\rm local \;\,  type:} \;\;\; \tfNLd(\bk_1,\bk_2)= \fNL \;
\frac{\alpha(\bk_1+\bk_2)}{\alpha(k_1) \, \alpha(k_2)} .
\label{tfNLd_local}
\eeq
This reads in real space as
\beq
\delta_L(\bx) = \chi(\bx) + \int \dd\bx_1\dd\bx_2 \, \fNLd(\bx;\bx_1,\bx_2) 
\, \chi(\bx_1) \, \chi(\bx_2) ,
\label{fNLd_def}
\eeq
where the kernel $\fNLd(\bx;\bx_1,\bx_2)$ only depends on the two vectors
$\{\bx_1-\bx,\bx_2-\bx\}$, 
\beq
\fNLd(\bx;\bx_1,\bx_2) = \fNLd(\bx_1-\bx,\bx_2-\bx) ,
\label{fNL_homogeneous}
\eeq
as long as the system remains statistically homogeneous, as for the local model
(\ref{fNLdef}).  The real-space and Fourier-space kernels are related by
(note the different normalization from (\ref{Fourier_norm}))
\beq
\fNLd(\bx_1,\bx_2) = \int \frac{\dd\bk_1\dd\bk_2}{(2\pi)^6} \,
e^{\ii \bk_1.\bx_1+\ii \bk_2.\bx_2} \, \tfNLd(\bk_1,\bk_2)  .
\label{tfNLd_def}
\eeq
The relationships (\ref{deltak_chik}) and (\ref{fNLd_def}) describe any
homogeneous model where the Bardeen potential can be expressed as the sum
of linear and quadratic terms over some Gaussian field. Thus, our analytical
results also apply to other ``$\fNL$-type'' models than the ``local''
one (\ref{fNLdef}). 

For $\fNL=0$ we recover Gaussian initial conditions, $\delta_L=\chi$, with
a linear density power spectrum
\beq
\lag\tchi(\bk_1)\tchi(\bk_2)\rag = \delta_D(\bk_1+\bk_2) P_L(k_1) ,
\label{PLkdef}
\eeq
and a two-point linear density correlation
\beqa
C_L(\bx_1,\bx_2) & = & \lag \chi(\bx_1) \chi(\bx_2) \rag \nonumber \\
& = & 4\pi \int \dd k \, k^2 \, P_L(k) \, 
\frac{\sin(k|\bx_2-\bx_1|)}{k|\bx_2-\bx_1|} .
\label{CLdef}
\eeqa
As usual, it is convenient to introduce the  smoothed linear density contrast,
$\chi_q(\bx)$, within the sphere of radius $q$ and volume $V$ around position
$\bx$,
\beq
\chi_q(\bx) = \int_V \frac{\dd \bx'}{V} \, \chi(\bx+\bx') 
= \int\dd\bk \, e^{\ii\bk.\bx} \, \tchi(\bk) \, \tW(k q) ,
\label{chiqdef}
\eeq
with a top-hat window that reads in Fourier space as 
\beq
\tW(k q) = \int_V \frac{\dd \bx}{V} \, e^{\ii\bk.\bx} = 
3 \, \frac{\sin(k q)-k q\cos(k q)}{(k q)^3} .
\label{Wdef}
\eeq
Then, in the linear regime, the cross-correlation of the smoothed linear density
contrasts on scales $q_1$ and $q_2$ and positions $\bx_1$ and
$\bx_2=\bx_1+\bx$ reads as
\beqa
\sigma^2_{q_1,q_2}(x) & = & \lag \chi_{q_1}(\bx_1)
\chi_{q_2}(\bx_1+\bx) \rag  \nonumber \\
& = & 4\pi\int\dd k \, k^2 P_L(k) \tW(k q_1) \tW(k q_2) 
\frac{\sin(kx)}{kx} .
\label{sigq1q2}
\eeqa
In particular, $\sigma_q=\sigma_{q,q}(0)$ is the usual rms linear density
contrast on scale $q$.
Then, in the non-Gaussian case we define the initial conditions by the same
power spectrum (\ref{PLkdef}) for the field $\chi$ and we vary the
parameter $\fNL$ (for the local type (\ref{tfNLd_local})). We still
define the variance $\sigma^2$ as in Eq.(\ref{sigq1q2}) from the Gaussian
field $\chi$.

In the following sections, where we define dark matter halos as spherical
overdensities, we shall need the average of the kernel $\fNLd(\bx;\bx_1,\bx_2)$
over spherical cells, weighted by the linear correlation (\ref{CLdef}).
Thus, omitting the superscript $\delta$ for simplicity, we define the quantity
\beqa
f_{q;q_1,q_2}(\bx;\bx_1,\bx_2) & = & \int_V\frac{\dd\bq}{V} 
\int_{V_1}\frac{\dd\bq_1}{V_1} \int_{V_2}\frac{\dd\bq_2}{V_2} 
\int \dd\bq_1'\dd\bq_2'  \nonumber \\
&& \hspace{-1cm} \times \,\fNLd(\bq;\bq_1',\bq_2') \, 
C_L(\bq_1',\bq_1) \, C_L(\bq_2',\bq_2) ,
\label{fq3def}
\eeqa
where the spheres of volumes $V$, $V_1$ and $V_2$, and radii $q$, $q_1$ and
$q_2$, are centered on the points $\bx$, $\bx_1$ and $\bx_2$. 
In Eq.(\ref{fq3def}) the coordinates $\bq_1'$ and $\bq_2'$ are integrated over
all space. In terms of the Fourier kernel $\tfNLd$ this reads as
\beqa
f_{q;q_1,q_2}(\bx;\bx_1,\bx_2) & = & \int \dd\bk_1\dd\bk_2 \, 
\tfNLd(\bk_1,\bk_2) \, P_L(k_1) P_L(k_2) \nonumber \\
&& \times \, \tW(k_1 q_1) \, \tW(k_2 q_2) \,
\tW(|\bk_1+\bk_2|q) \nonumber \\
&& \times \, e^{\ii\bk_1.(\bx_1-\bx)+\ii\bk_2.(\bx_2-\bx)} .
\label{fq3_Four}
\eeqa
Thanks to statistical homogeneity and isotropy, the kernel
$\tfNLd(\bk_1,\bk_2)$ only depends on the lengths $k_1$, $k_2$, and on the
angle between both vectors. Then, for spheres that are centered on the
same point (i.e. $\bx=\bx_1=\bx_2$), Eq.(\ref{fq3_Four}) simplifies as
\beqa
\bx \! = \! \bx_1 \! = \! \bx_2 : \;\;\; f_{q;q_1,q_2} & = & 8\pi^2 
\int_0^{\infty} \dd k_1 \, k_1^2 P_L(k_1) \tW(k_1 q_1) \nonumber \\
&& \hspace{-3.6cm} \times \int_0^{\infty} \dd k_2 \, k_2^2 P_L(k_2) \tW(k_2 q_2)
\int_{-1}^1 \dd\mu \, \tW(k q) \tfNLd(k_1,k_2,\mu) ,  \nonumber \\ 
\label{fq3_x0}
\eeqa
which does not depend on the position $\bx$ of the sphere. Here we defined
$\mu=(\bk_1.\bk_2)/(k_1k_2)$, and $k^2=k_1^2+k_2^2+2k_1k_2\mu$.
On the other hand, when we consider the two-point correlation of dark matter
halos, the three spheres in Eq.(\ref{fq3def}) are chosen among two possible
spheres $V_a$ and $V_b$, separated by a distance $x$. 
Then, we need the two quantities,
\beqa
V=V_a, \; V_1=V_2=V_b: \;\; f_{a;bb}(x) & = & 8 \pi^2  \int_0^{\infty} 
\dd k_1 \, k_1^2 P_L(k_1) \nonumber \\
&& \hspace{-4cm} \times \tW(k_1 q_b) \int_0^{\infty} \dd k_2 \, k_2^2 P_L(k_2)
\tW(k_2 q_b) \nonumber \\ 
&& \hspace{-4cm} \times \int_{-1}^1 \dd\mu \, \tW(k q_a) \tfNLd(k_1,k_2,\mu)
\frac{\sin(k x)}{k x} ,
\label{fabb}
\eeqa
and
\beqa
V_1=V_a, \; V=V_2=V_b: \;\; f_{b;ab}(x) & = & 8 \pi^2  \int_0^{\infty} 
\dd k_1 \, k_1^2 P_L(k_1) \nonumber \\
&& \hspace{-4cm} \times \tW(k_1 q_a) \frac{\sin(k_1 x)}{k_1 x} 
\int_0^{\infty} \dd k_2 \, k_2^2 P_L(k_2) \tW(k_2 q_b) \nonumber \\ 
&& \hspace{-4cm} \times \int_{-1}^1 \dd\mu \, \tW(k q_b) \tfNLd(k_1,k_2,\mu) .
\label{fbab}
\eeqa
Their Fourier transforms with respect to the separation $\bx$ read as
\beqa
\tf_{a;bb}(k) & = & \int \dd \bk_1 \dd \bk_2 \, \delta_D(\bk_1+\bk_2-\bk)
P_L(k_1) P_L(k_2) \nonumber \\
&& \times \tW(k_1 q_b) \tW(k_2 q_b) \tW(k q_a) \tfNLd(\bk_1,\bk_2) ,
\label{tfabb}
\eeqa
and
\beqa
\tf_{b;ab}(k) & = & P_L(k) \tW(k q_a) \int \dd \bk_1 \, P_L(k_1) \tW(k_1 q_b) 
 \nonumber \\
&& \times \tW(|\bk_1+\bk| q_b) \tfNLd(\bk,\bk_1) .
\label{tfbab}
\eeqa

\section{Mass function of dark matter halos}
\label{Mass-function-halos}

We now extend the analysis of Valageas (2009b) to obtain the mass function
of dark matter halos for non-Gaussian initial conditions.

\subsection{Rare-event saddle point for the density distribution}
\label{saddle-point}

In a fashion similar to Valageas (2009b), we note that the exponential falloffs
of the high-mass tail of the halo mass function $n(M)$, and of the
overdensity tail of the linear density contrast distribution $\cP_L(\delta_{Lq})$,
can be exactly obtained from the constrained maximum,
\beq
\mbox{rare events}: \;\;  \cP_L(\delta_L) \sim \max_{\{\chi[\bq]{\displaystyle |}
\delta_{Lq}=\delta_L\}} e^{-\frac{1}{2} \chi.C_L^{-1}.\chi} .
\label{rareP}
\eeq
Here we introduced the probability distribution $\cP_L(\delta_{Lq})$ of the
smoothed linear density contrast within the sphere of radius $q$, which
we can take centered on the origin,
\beq
\delta_{Lq} = \int_V \frac{\dd \bq'}{V} \, \delta_L(\bq') 
= \int\dd\bk \, \tdelta_L(\bk) \, \tW(kq) .
\label{deltaLqdef}
\eeq
In the Gaussian case, $\fNL=0$, we simply have $\delta_{Lq}=\chi_q$, as defined
in Eq.(\ref{chiqdef}), and the probability distribution $\cP_L$ is a Gaussian.
Then, as stressed in Valageas (2009a,b), in the limit of rare events (e.g., at
fixed density contrast $\delta_{Lq}$ in the large-scale or high-mass limit
$q\rightarrow\infty$), the tails of the distribution $\cP_L$ are obtained
from Eq.(\ref{rareP}), where we maximize the statistical weight
$e^{-(\chi.C_L^{-1}.\chi)/2}$
of the Gaussian field $\chi$, under the constraint that the smoothed linear
density contrast $\delta_{Lq}$ is equal to the value of interest.
Equation (\ref{rareP}) only gives the leading-order exponential falloff.
Subleading terms, such as power-law prefactors, may be obtained by expanding
around the saddle-point $\chi$, within a steepest-descent method. This also
provides the probability distribution $\cP(\delta)$ of the nonlinear density contrast
$\delta$ on scale $r$ in the quasi-linear regime, as well as the cumulant generating
function (Valageas 2002a,b; 2009a,b)\footnote{As shown in Valageas (2009a),
for the closely related adhesion model, where the same procedure can be
applied, one can explicitly check that the asymptotic results obtained by this approach
agree with the complete distribution $\cP(\delta)$ that is exactly known for
two cases (Brownian and white-noise linear velocity in 1-D, corresponding
to a power-law linear density power spectrum with $n=-2$ and $n=0$).}. 
As pointed out in Valageas (2009b), this also gives the high-mass tail of the
halo mass function, if halos are defined as spherical overdensities with
a fixed nonlinear density threshold $\delta_r$. Then, the halo mass $M$ and radius
$r$ are related to the Lagrangian radius $q$ and linear density threshold
$\delta_{Lq}$ through
\beq
q^3 = (1+\delta_r) \, r^3 , \;\;\; \mbox{with} \;\;\;
\delta_{r} = \cF(\delta_{Lq}) ,
\label{qr}
\eeq
and
\beq
M= \rhobm \frac{4\pi}{3} q^3 ,
\label{Mdef}
\eeq
where the function $\cF$ describes the spherical collapse dynamics. 
This holds as long as shell-crossing has not extended beyond radius $r$,
so that the usual spherical collapse dynamics at constant mass is valid.
This yields an upper bound $\delta_+$ for the nonlinear density threshold
that can be used to define halos to take advantage of the exact asymptotic
tail (\ref{rareP}). For the preferred $\Lambda$CDM model this gives
$\delta_+ \sim 200$ for $M\sim 10^{15}h^{-1}M_{\odot}$ up to 
$\delta_+ \sim 600$ for $M\sim 10^{11}h^{-1}M_{\odot}$
(the dependence on mass is due to the change of slope of the matter power
spectrum $P_L(k)$).

In the Gaussian case, the constrained weight (\ref{rareP}) simply reads as
$e^{-\delta_{Lq}^2/(2\sigma_q^2)}$, and we recover the exponential tail of the
usual Press-Schechter mass function (Press \& Schechter 1974), except that
the standard threshold $\delta_c\simeq 1.686$ must be replaced by
$\delta_L=\cF^{-1}(\delta)$, with $\delta_L\simeq 1.59$ for $\delta=200$.
In the non-Gaussian case, that is for $\fNL\neq 0$, we can compute
the weight (\ref{rareP}) by using a Lagrange multiplier $\lambda$.
Thus, we define the action $\cS[\chi,\lambda]$,
\beq
\cS[\chi,\lambda] = \lambda \, \left(\delta_L - \delta_{Lq}[\chi]\right) 
+ \frac{1}{2} \, \chi.C_L^{-1}.\chi
\label{Sdef}
\eeq
where $\delta_{Lq}[\chi]$ is the nonlinear functional that affects to the initial
condition defined by the Gaussian field $\chi$ the linear density contrast
$\delta_{Lq}$ within the sphere of radius $q$, obtained through
Eq.(\ref{fNLd_def}). Then, we must look for the saddle point of the action
(\ref{Sdef}), with respect to both $\chi$ and $\lambda$.
Thus, differentiating the action (\ref{Sdef}) with respect to $\chi(\bq)$
and multiplying by the operator $C_L$ gives
\beq
\chi(\bq) = \lambda \, C_L(\bq,\bq') . \frac{\cD \delta_{Lq}}{\cD \chi(\bq')} ,
\eeq
whence, using Eq.(\ref{fNLd_def}),
\beqa
\chi(\bq) & = & \lambda \int_V\frac{\dd\bq'}{V} \, C_L(\bq,\bq') 
+ 2 \lambda \int_V \frac{\dd\bq'}{V} \int\dd\bq_1\dd\bq_2 \nonumber \\
&& \times \, \fNLd(\bq';\bq_1,\bq_2) \, C_L(\bq,\bq_2) \, \chi(\bq_1) .
\label{dSdchi}
\eeqa
Differentiating the action (\ref{Sdef}) with respect to $\lambda$
gives the constraint
\beq
\delta_L = \delta_{Lq}[\chi] ,
\label{constraint}
\eeq
whence
\beqa
\delta_L \! & = & \!\! \int_V \! \frac{\dd\bq}{V} \left[ \chi(\bq) + \! \int \! 
\dd\bq_1\dd\bq_2 \, \fNLd(\bq;\bq_1,\bq_2) \, \chi(\bq_1) \, \chi(\bq_2) 
\right] . \nonumber \\ &&
\label{constraintchi}
\eeqa
Next, we solve the system (\ref{dSdchi})-(\ref{constraintchi}) as a perturbative
series over the non-Gaussianity kernel $\fNLd$ (i.e. over powers of the
parameter $\fNL$). At order zero we recover the Gaussian saddle-point,
\beqa
\chi^{(0)}(\bq) & = & \lambda^{(0)} \int_V\frac{\dd\bq'}{V} \, C_L(\bq,\bq') ,
\label{chi0} \\
\lambda^{(0)} & = & \frac{\delta_L}{\sigma_q^2} .
\label{lambda0}
\eeqa
At first order we obtain
\beqa
\chi^{(1)}(\bq) & = & \lambda^{(1)} \int_V\frac{\dd\bq'}{V} \, C_L(\bq,\bq') 
+ 2 \lambda^{(0)2} \int_V \frac{\dd\bq'\dd\bq_1'}{V^2} \nonumber \\
&& \hspace{-1cm} \times \int\dd\bq_1\dd\bq_2 \, \fNLd(\bq';\bq_1,\bq_2) \, 
C_L(\bq,\bq_2) \, C_L(\bq_1,\bq_1') ,
\label{chi1}
\eeqa
\beq
\lambda^{(1)} = -3 \, \frac{\delta_L^2}{\sigma_q^6} \, f_{q;qq} ,
\label{lambda1}
\eeq
where the quantity $f_{q;qq}$ is given by Eq.(\ref{fq3_x0}).

\begin{figure}[htb]
\begin{center}
\epsfxsize=8.6 cm \epsfysize=6. cm {\epsfbox{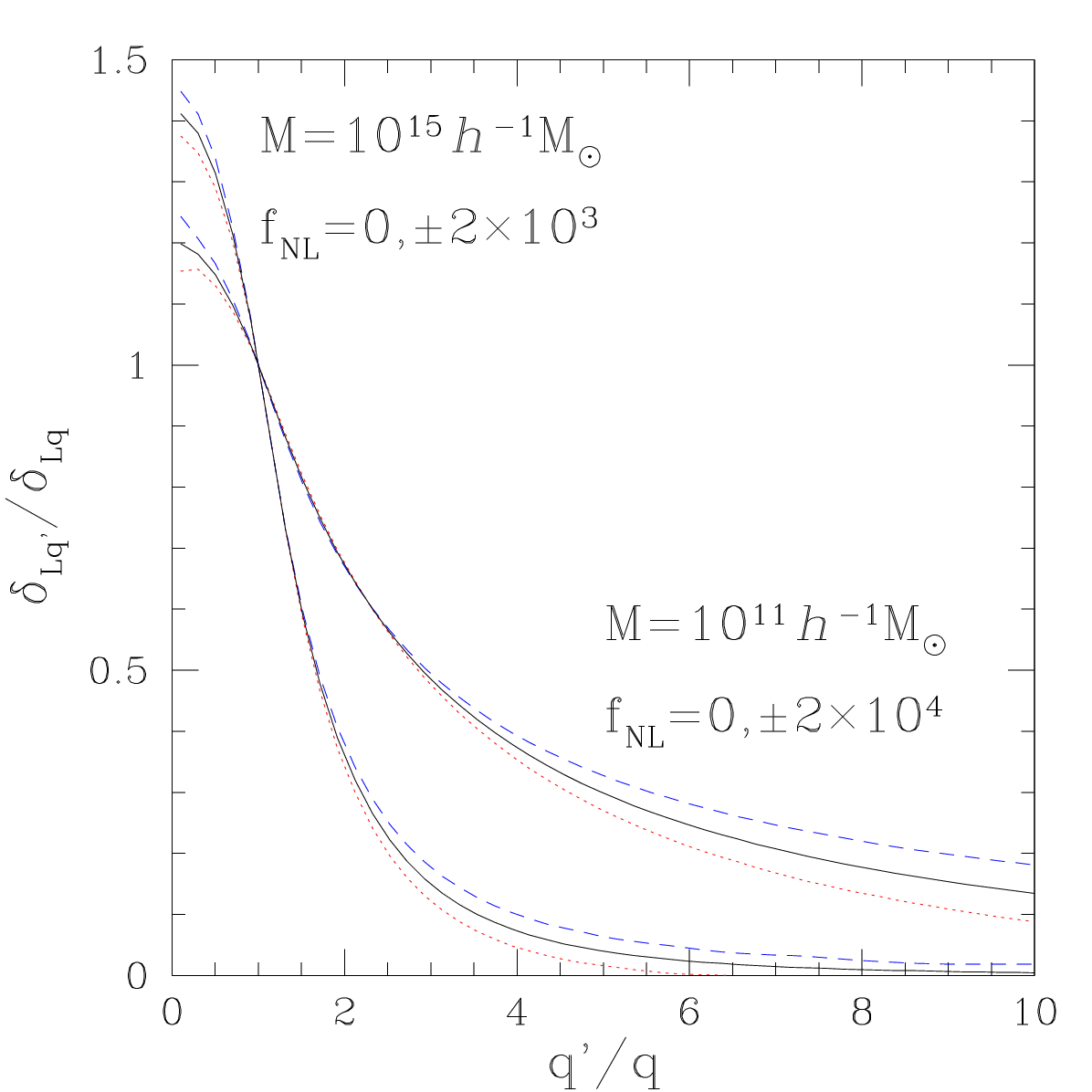}}
\end{center}
\caption{The radial profile (\ref{delta0})-(\ref{delta1}) of the linear
density contrast $\delta_{Lq'}$ of the saddle point of the action
$\cS[\chi,\lambda]$.
We show the profiles obtained with a $\Lambda$CDM cosmology for the
masses $M=10^{11}$ and $10^{15}h^{-1}M_{\odot}$.
A larger mass corresponds to a lower ratio $\delta_{Lq'}/\delta_{Lq}$
at large radii $q'/q>1$. We show our results for the Gaussian case (solid line),
positive $\fNL$ (dashed line) and negative $\fNL$ (dotted line), for the
local model (\ref{tfNLd_local}).}
\label{figdeltaLq}
\end{figure}

From Equations (\ref{chi0})-(\ref{lambda1}) we also obtain the radial linear
density profile of the saddle-point, up to first order,
\beqa
\delta_{Lq'}^{(0)} & = & \frac{\delta_L}{\sigma_q^2} \, \sigma_{q,q'}^2 ,
\label{delta0} \\
\delta_{Lq'}^{(1)} & = & \frac{\delta_L^2}{\sigma_q^4} \, 
\left[ f_{q';qq} + 2 f_{q;qq'} - 3 \frac{\sigma_{q,q'}^2}{\sigma_q^2} f_{q;qq}
\right] .
\label{delta1} 
\eeqa
We can check that at $q'=q$ it verifies the constraint (\ref{constraint}), and
at order zero we recover the Gaussian profile (Valageas 2009b).
Equations (\ref{delta0})-(\ref{delta1}) give the integrated density profile,
that is, $\delta_{Lq'}$ is the mean linear density contrast within the
Lagrangian radius $q'$. The local linear density contrast at radius $q'$,
$\delta_L(q')$, is given by
\beq
3 q'^3 \delta_L(q') = \frac{\pl}{\pl q'} (q'^3 \delta_{Lq'}) ,
\eeq
whence
\beqa
\delta_L^{(0)}(q') & = & \frac{\delta_L}{\sigma_q^2} \, \sigma^2_{q,0}(q') ,
\label{deltap0} \\
\delta_L^{(1)}(q') & = & \frac{\delta_L^2}{\sigma_q^4} \, 
\left[ f_{0;qq}(q') + 2 f_{q;0q}(q') - 3 \frac{\sigma^2_{q,0}(q')}{\sigma_q^2}
f_{q;qq} \right] 
\label{deltap1} 
\eeqa
where $f_{0;qq}(q')$ and $f_{q;0q}(q')$ are given by
Eqs.(\ref{fabb})-(\ref{fbab}).

We show in Fig.~\ref{figdeltaLq} the integrated linear density profile
(\ref{delta0})-(\ref{delta1}) obtained for the masses $M=10^{11}$ and
$10^{15}h^{-1}M_{\odot}$, for a $\Lambda$CDM cosmology. The dependence on
mass is due to the change of slope of the linear power spectrum with scale.
We plot our results for the Gaussian case ($\fNL=0$, solid lines), large
positive $\fNL$ ($\fNL=2\times 10^{3}$ and $\fNL=2\times 10^{4}$, dashed lines)
and large negative $\fNL$ ($\fNL=-2\times 10^{3}$ and $\fNL=-2\times 10^{4}$,
dotted lines), for the local model (\ref{tfNLd_local}).
Thus, we can see that a positive $\fNL$ increases the relative
density (i.e. with respect to the density at radius $q$) both at small and large radii.
The very large values of $\fNL$ required to be able to distinguish the curves in the
figure imply that for realistic cases ($|\fNL|<100$) the perturbation of the
density profile is very small. Therefore, the values of the upper boundary
$\delta_+$, which marks the onset of shell-crossing, obtained in
Valageas (2009b) for the Gaussian case remain valid up to a very good accuracy.
We can note that to obtain a similar deviation from the Gaussian profile
we need a larger value of the parameter $\fNL$ on a smaller scale.
This can be understood from the expression (\ref{tfNLd_local}), which scales
as $\tfdel \sim \fNL /\alpha(k) \propto \fNL/(k^2 T(k))$ and grows as $k^{-2}$
on very large scales. The same behavior (i.e. a higher sensitivity to local-type
non-Gaussianity on large scales) is obtained for the bias of dark matter halos,
see Eq.(\ref{Dbk_Dal}) above and section~\ref{Bias-of-halos} below.

Next, we define the constrained weight (\ref{rareP}) as
\beq
\Gam = \frac{1}{2} \chi . C_L^{-1} . \chi
\label{Gamdef}
\eeq
at the relevant saddle-point $\chi$. This gives, up to first order,
\beqa
\Gam^{(0)} & = & \frac{1}{2} \chi^{(0)} . C_L^{-1} . \chi^{(0)} = 
\frac{\delta_L^2}{2\sigma_q^2} , \label{Gam0} \\
\Gam^{(1)} & = & \chi^{(0)} . C_L^{-1} . \chi^{(1)} 
= - \, \frac{\delta_L^3}{\sigma_q^6} \, f_{q;qq} . \label{Gam1}
\eeqa
Therefore, the tails of the probability distribution (\ref{rareP}) read
as
\beq
\cP_L(\delta_L) \sim e^{-\frac{\delta_L^2}{2\sigma_q^2}
+ \frac{\delta_L^3}{\sigma_q^6} f_{q;qq}}
= e^{-\frac{\delta_L^2}{2\sigma_q^2} 
\left( 1-\frac{\delta_L}{3} S_3^{(1)}\right)} ,
\label{PS3}
\eeq
where we introduced the skewness of the linear density contrast, at
first order over $\fNL$, 
\beq
S_3^{(1)} = \left[ \frac{\lag\delta_{Lq}^3\rag}{\lag\delta_{Lq}^2\rag^2}
\right]^{(1)} = 6 \, \frac{f_{q;qq}}{\sigma_q^4} .
\label{S3def}
\eeq
Indeed, from Eq.(\ref{fNLd_def}) we have at first order
\beqa
\lag\delta_{Lq}^3\rag^{(1)} & = & 3 \int_V\frac{\dd\bq_1\dd\bq_2\dd\bq_3}{V^3}
\int\dd\bq_1'\dd\bq_2' \, \fNLd(\bq_3;\bq_1'\bq_2') \nonumber \\
&& \times \lag \chi(\bq_1) \chi(\bq_2) \chi(\bq_1') \chi(\bq_2') \rag \\
& = & 6 \, f_{q;qq} .
\eeqa
Hereafter, since $S_3^{(0)}=0$, we simply note $S_3=S_3^{(1)}$.

\subsection{Mass function}
\label{Mass-function}

Following the Press-Schechter approach (Press \& Schechter 1974), the mass
function that is obtained in the Gaussian case from the probability distribution
$\cP_L(\delta_{Lq})$ reads as
\beq
n(M,\fNL=0) \dd M = \frac{\rhobm}{M} f(\nu) \frac{\dd\nu}{\nu} ,
\label{massF}
\eeq
with (using the subscript ``PS'' to distinguish the Press-Schechter prediction)
\beq
f_{\rm PS}(\nu) = \sqrt{\frac{2}{\pi}} \, \nu \, e^{-\nu^2/2}
\label{fPS}
\eeq
and
\beq
\nu = \frac{\delta_L}{\sigma(M)} .
\label{nudef}
\eeq
Here $\sigma(M)=\sigma_q$, where the Lagrangian scale $q$ is related to $M$ by
Eq.(\ref{Mdef}). As stressed in Valageas (2009b), the linear threshold
$\delta_L$ in Eq.(\ref{nudef}) must be defined as $\cF^{-1}(\delta)$, as in
Eq.(\ref{qr}), which gives $\delta_L \simeq 1.59$ for $\delta=200$.
Then, the exponential tail of (\ref{fPS}) is exact, 
\beq
\nu \rightarrow \infty : \;\; \ln[f(\nu)] \sim -\frac{\nu^2}{2} ,
\label{nutail}
\eeq
but the power-law prefactor and the low-mass tail of (\ref{fPS}) have no reason to
be valid (and numerical simulations indeed show that they are not exact).
Then, in order to match numerical
simulations, one needs to use fitting formulae. One such fit to simulations,
which obeys the exact tail (\ref{nutail}), is (Valageas 2009b)
\beq
f(\nu) = 0.5 \left[ (0.6 \, \nu)^{2.5}+(0.62 \, \nu)^{0.5} \right] 
\, e^{-\nu^2/2} .
\label{fitfsig}
\eeq
Both mass functions (\ref{fPS}) and (\ref{fitfsig}) satisfy the normalization
\beq
\int_0^{\infty} \frac{\dd\nu}{\nu} \, f(\nu) = 1 ,
\label{normf}
\eeq
which ensures that all the mass is contained in such halos:
\beq
\int_0^{\infty}  M \, n(M) \dd M = \rhobm .
\label{norm_n}
\eeq

In the non-Gaussian case, that is $\fNL\neq 0$, we may estimate the halo mass
function by multiplying the Gaussian one by the corrective factor obtained in
Eq.(\ref{PS3}),
\beq
\fNL\neq 0: \;\; n(M,\fNL) = n(M,0) \, e^{S_3 \delta_L^3 /(6\sigma_q^2)} .
\label{mult}
\eeq
As explained above, this yields the exact high-mass tail (up to first order
over $\fNL$) but it is not expected to hold for the low-mass tail.
In particular, this gives a non-Gaussian mass function that does not obey the
normalization (\ref{norm_n}). In order to satisfy Eq.(\ref{norm_n}), while
keeping the high-mass tail of Eq.(\ref{mult}), a simple procedure is to
make use of the scaling (\ref{massF}) and of the normalization (\ref{normf}).
Thus, modifying the relationship (\ref{nudef}) as (see Eq.(\ref{PS3}))
\beq
\mu = \frac{\delta_L}{\sigma(M)} \, \sqrt{1-\frac{\delta_L}{3}S_3(M)} ,
\label{muS3def}
\eeq
we may use for the non-Gaussian mass function
\beq
\fNL\neq 0: \;\;  n(M,\fNL) \dd M = \frac{\rhobm}{M} f(\mu) \frac{\dd\mu}{\mu} ,
\label{massF-fNL}
\eeq
where we use the same scaling function $f$ as for the Gaussian case
(\ref{massF}). This recovers the high-mass tail (\ref{mult}) and satisfies the
normalization (\ref{norm_n}).
Note that Afshordi \& Tolley (2008) have recently proposed a similar rescaling
as (\ref{muS3def}), which they re-interprate as a modified effective variance
$\sigma_q^2$. However, to obtain such a rescaling they use some approximations,
such as the decoupling between the local values of the fields $\phi(\bx)$ and
$\chi(\bx)$, in the spirit of a peak-background split approximation, whereas
the derivation presented here is asymptotically exact. 
Equation (\ref{massF-fNL}) yields for the ratio of both mass functions at fixed
mass $M$,
\beq
\frac{n(M,\fNL)}{n(M,0)} = \frac{f(\mu)/\mu}{f(\nu)/\nu} \, \frac{\dd\mu}{\dd\nu} .
\label{rationM}
\eeq
Using Eqs.(\ref{nudef}), (\ref{muS3def}), we obtain
\beq
\frac{\dd\mu}{\dd\nu} = \sqrt{1-\frac{\delta_L}{3}S_3} 
+ \frac{\sigma \, \delta_L}{6\sqrt{1-\frac{\delta_L}{3}S_3}} 
\, \frac{\dd S_3}{\dd\sigma} .
\label{munu}
\eeq
If we use the Press-Schechter mass function (\ref{fPS}), 
Eqs.(\ref{massF-fNL})-(\ref{munu})
give back the result obtained in Matarrese et al. (2000), except for the fact
that we use $\delta_L\simeq 1.59$ instead of $1.686$ as explained above.

Note that Matarrese et al. (2000) also use a saddle-point approach to
derive the halo mass function for non-Gaussian initial conditions, expanding
at linear order over $\fNL$ (or $S_3$), so that their computation is equivalent
to the one described above. However, since they first compute the cumulant
generating function ($W(\lambda)$ in their notations, or $\varphi(y)$ in
Valageas 2009b), the constraint (\ref{constraint}) is expressed through
a Dirac function, written as an exponential by introducing an auxiliary variable
$\lambda$, so that they can first integrate over the Gaussian field $\chi$
and next expand over $\fNL$ the expression obtained for $W(\lambda)$.
The somewhat simpler method described in this article has the advantage of
simultaneously giving the density profile (\ref{delta0})-(\ref{delta1}) of the
underlying saddle point. As explained above in Fig.~\ref{figdeltaLq}, this
allows us to check that realistic amounts of primordial non-Gaussianity
have a negligible effect on this profile, so that the onset of shell-crossing
appears for almost the same nonlinear density threshold $\delta_+$.
This ensures that the rare-event and high-mass tails (\ref{PS3}) and
(\ref{mult}) are exact (at leading order), as long as halos are defined by
a nonlinear threshold $\delta \sim 200$ below the upper bound $\delta_+$.
The simplicity of the method presented in this article also allows a
straightforward application to two-point distributions, as shown in
section~\ref{Bias-of-halos} below, or to more complex primordial
non-Gaussianities, which may involve several fields or higher order polynomials
as in Eq.(\ref{chi_i}) below.

Another approach presented in Lo Verde et al. (2008) is to use the Edgeworth
expansion, which writes the probability distribution $\cP_L(\delta_L)$ as a series
over the cumulants of the non-Gaussian variable $\delta_L$. In practice, one
truncates at the lowest order beyond the Gaussian, that is, at the third
cumulant $\lag\delta_L^3\rag_c$ described by $S_3$. Thus, expanding the
exponentials (\ref{PS3}) and (\ref{mult}) one recovers the results of
Lo Verde et al. (2008) at large $\nu$ (i.e. low $\sigma_q$). However,
in the rare-event limit, where computations rest on firm grounds as explained
above, the Edgeworth expansion does not fare very well.
In particular, although we only derived the saddle-point $\chi$ and the argument
$\Gamma$ of the exponential up to linear order over $\fNL$, see
Eqs.(\ref{Gam0})-(\ref{Gam1}), it is best to keep the exponential as in
Eqs.(\ref{mult}) or (\ref{massF-fNL}). Indeed, in the rare-event and small-$\fNL$
limits, the tail (\ref{mult}) can be a good approximation even when the corrective
factor $S_3\delta_L^3 /(6\sigma_q^2)$ is much larger than unity (i.e. we can
have the hierarchy $\Gamma^{(0)} \gg \Gamma^{(1)} \gg 1 \gg \Gamma^{(2)}$).

Since the low-mass tail (\ref{fPS}) does not match numerical simulations,
it is sometimes proposed to keep the ratio (\ref{rationM}) given by the
Press-Schechter mass function, and to multiply the fit from simulations of the
Gaussian mass function
by this factor (Grossi et al. 2007, 2009; Lo Verde et al. 2008). 
However, this procedure clearly violates
the normalization condition (\ref{norm_n}). Therefore, we suggest to use
(\ref{rationM}) with the fitting formula obtained from Gaussian simulations, that is,
to use Eq.(\ref{massF-fNL}), which automatically satisfies the normalization
(\ref{norm_n}). However, there is no reason to expect that the low-mass tail
can be exactly recovered by any such procedure, even though by construction
it gives the right behavior for the Gaussian case, as the low-mass slope might
also depend on $\fNL$ in some specific manner.

\begin{figure}[htb]
\begin{center}
\epsfxsize=4.44 cm \epsfysize=4.6 cm {\epsfbox{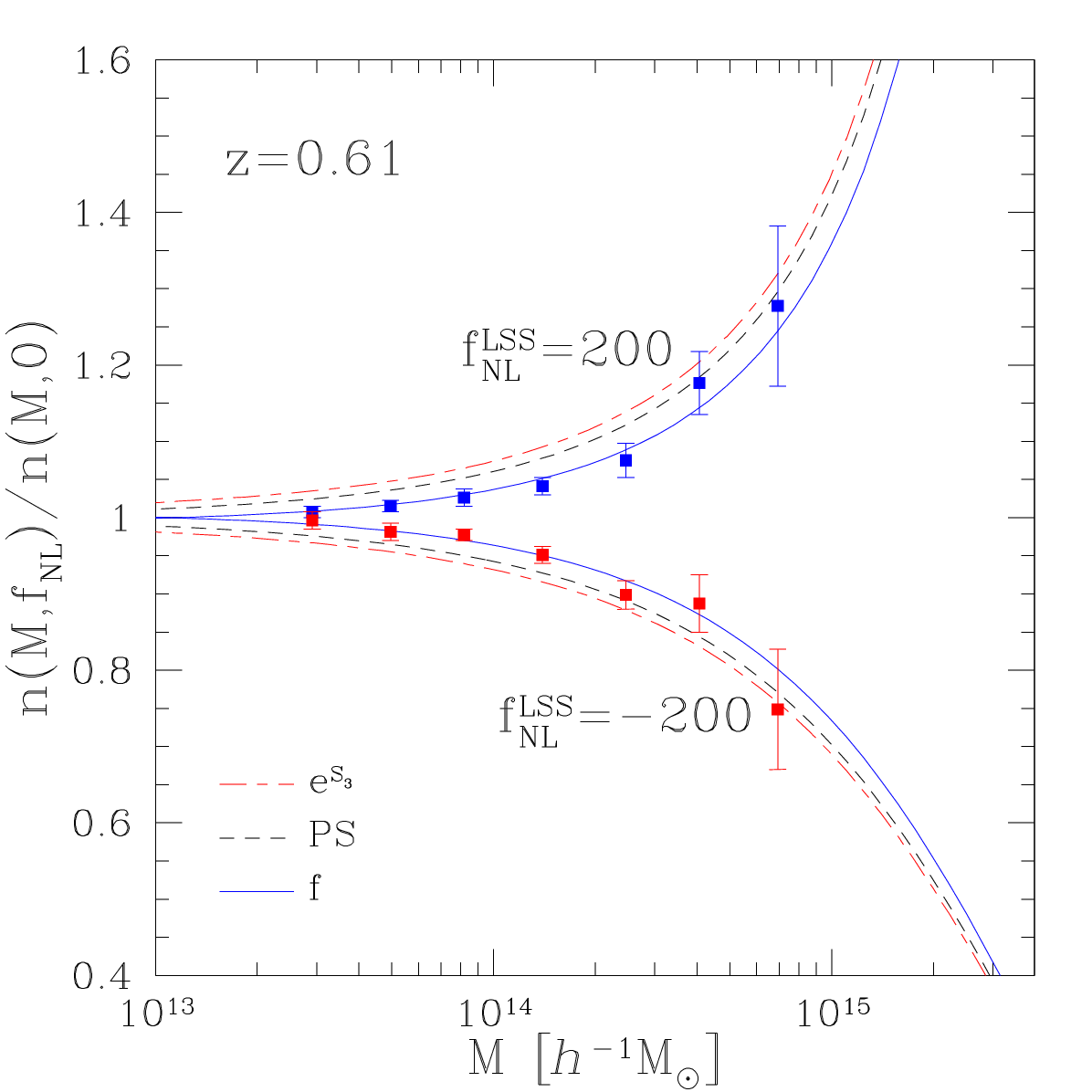}}
\epsfxsize=4.44 cm \epsfysize=4.6 cm {\epsfbox{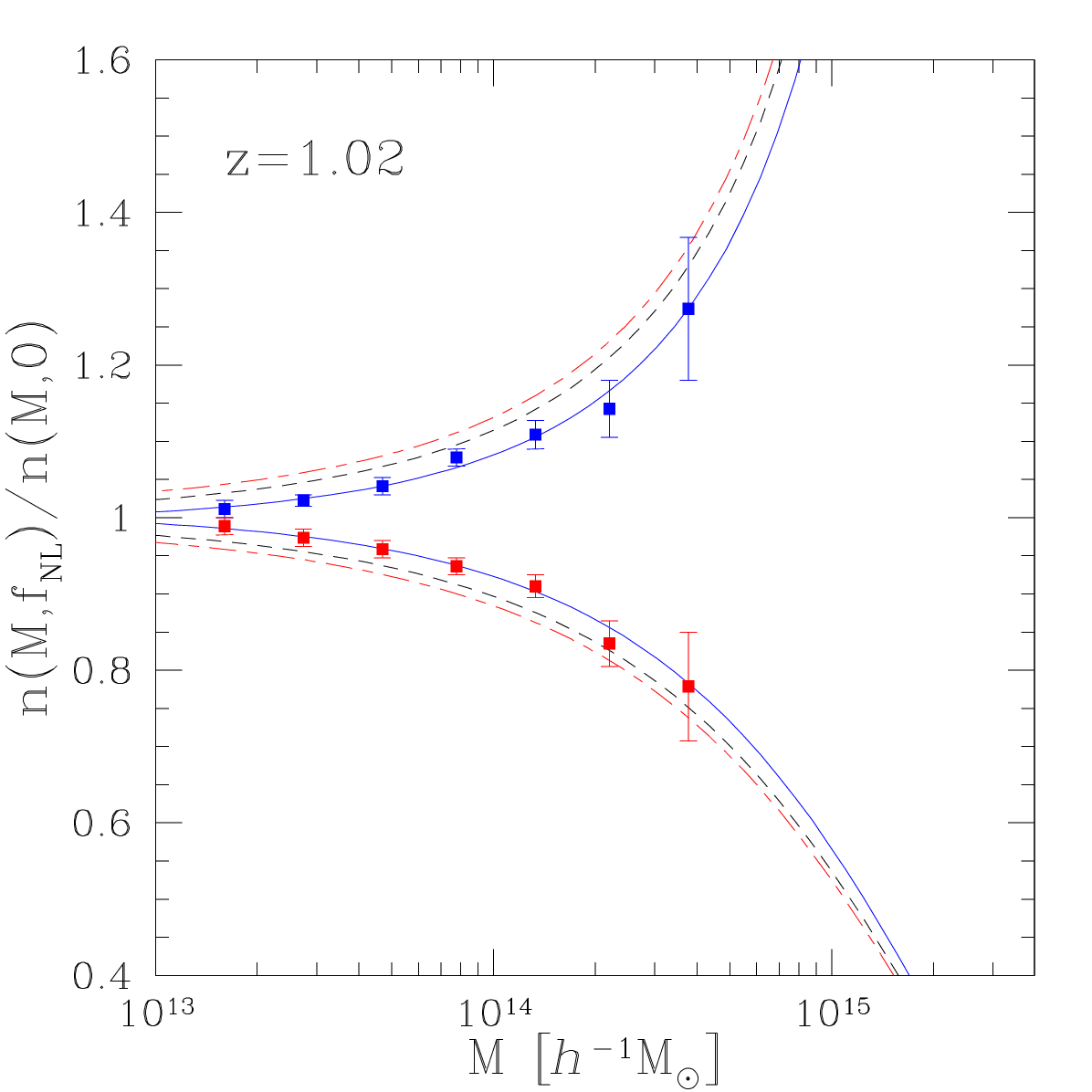}}\\
\epsfxsize=4.44 cm \epsfysize=4.6 cm {\epsfbox{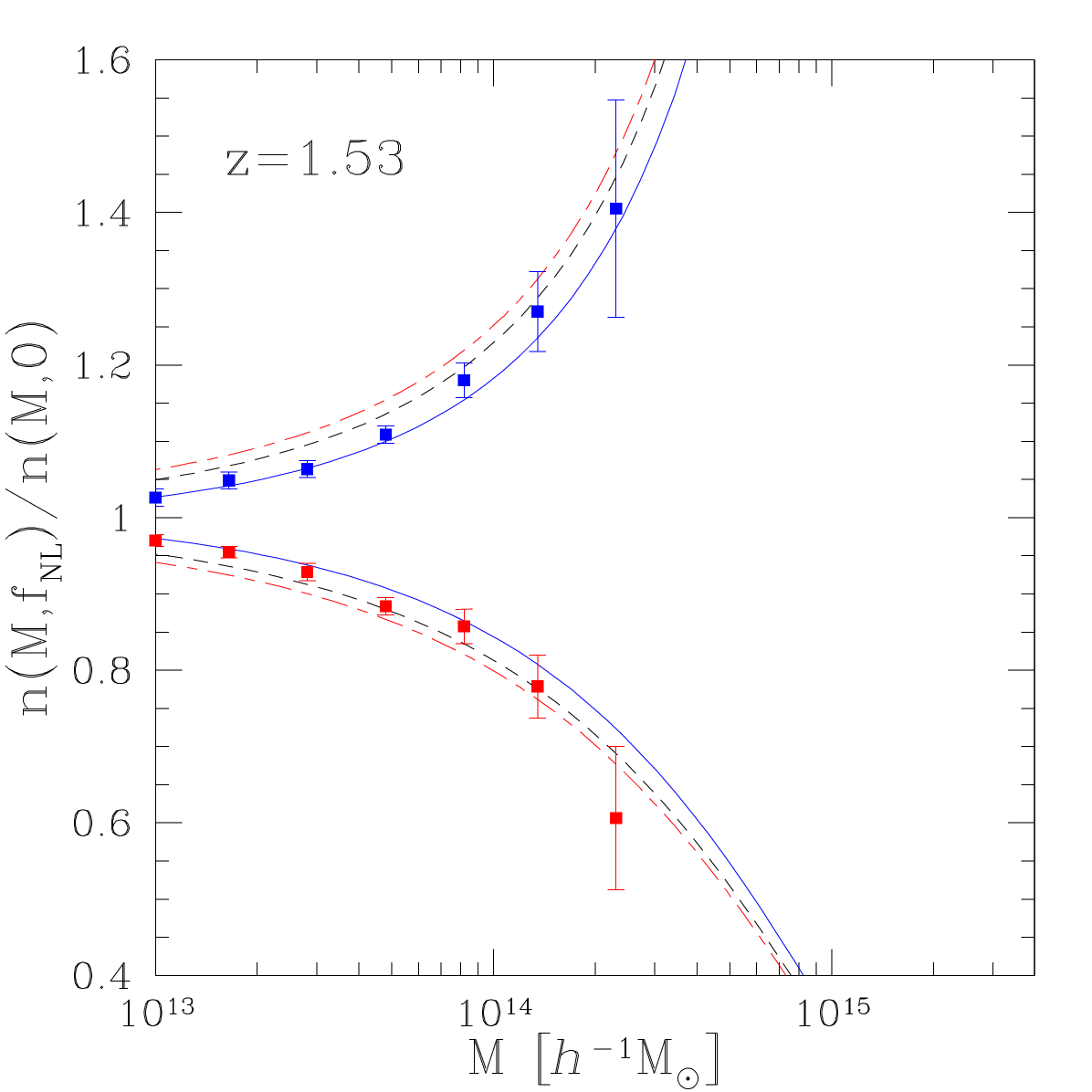}}
\epsfxsize=4.44 cm \epsfysize=4.6 cm {\epsfbox{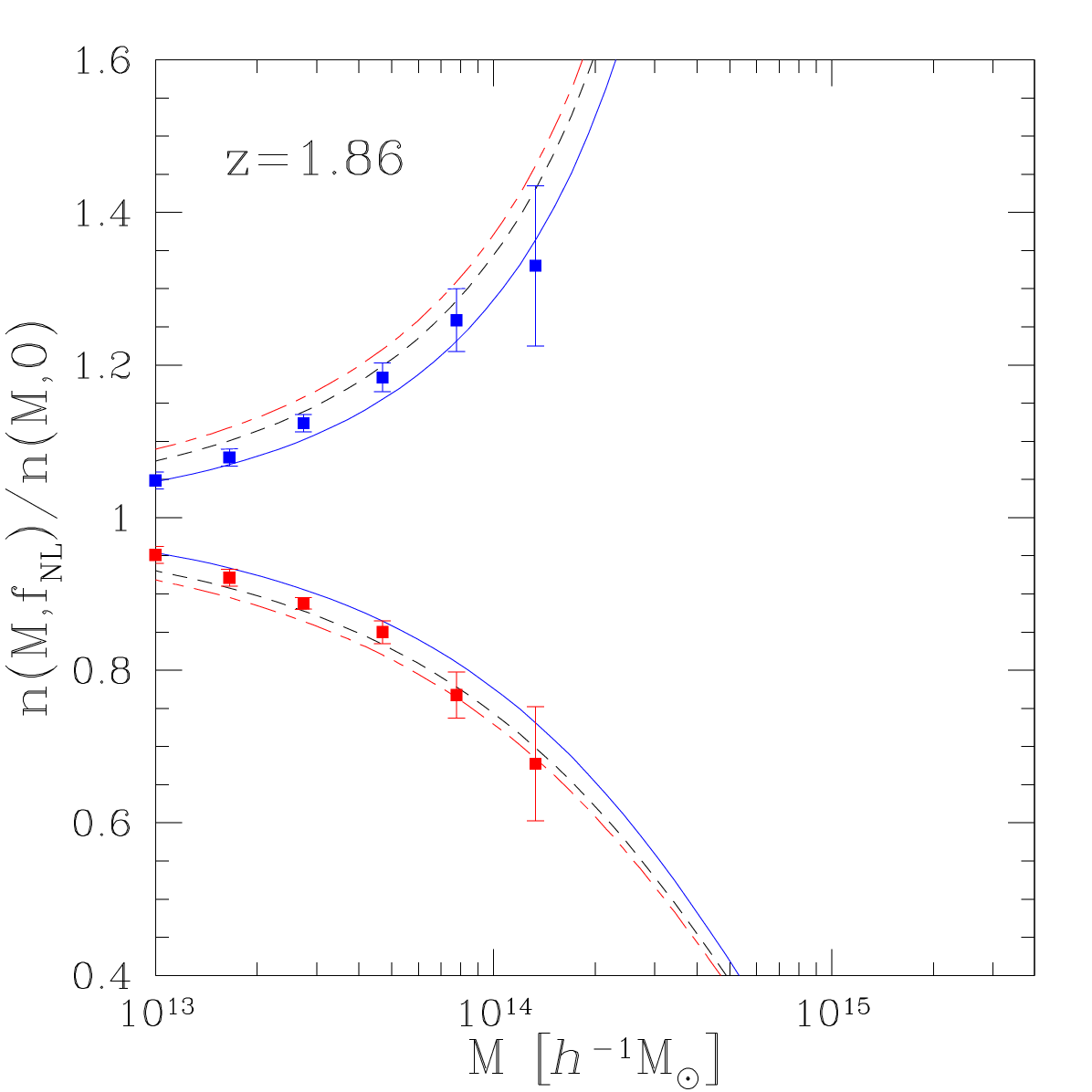}}
\end{center}
\caption{The ratio $n(M,\fNL)/n(M,0)$, of the mass functions obtained for
$\fNL^{\rm LSS}=\pm 200$ over the mass function obtained for Gaussian
initial conditions,
as a function of $M$ for several redshifts. The dot-dashed line labeled ``$e^{S_3}$''
is the multiplicative factor (\ref{mult}), the dashed line labeled ``PS'' is
Eq.(\ref{rationM}) with the Press-Schechter mass function (\ref{PS3}), which
also corresponds to the result of Matarrese et al. (2000)
(but with $\delta_L\simeq 1.59$), the solid line labeled ``f''
is Eq.(\ref{rationM}) with the fitting function (\ref{fitfsig}). The data points
are results from the numerical simulations of Grossi et al. (2009).}
\label{figfM_200_Grossi}
\end{figure}

We compare in Fig.~\ref{figfM_200_Grossi} our results
for the halo mass function with numerical simulations from Grossi et al. (2009).
They use the LSS convention for $\fNL$, so that in terms of the
CMB convention used in this article this corresponds to $\fNL \simeq \pm 151$. 
We show the ratio of the non-Gaussian mass function to the Gaussian one,
as given by the multiplicative factor (\ref{mult}), the ratio
(\ref{rationM}) computed with the Press-Schechter mass function (\ref{PS3})
or with the fitting function (\ref{fitfsig}).
In agreement with Grossi et al. (2009), we find that the ratio (\ref{rationM})
computed with the Press-Schechter mass function, which also corresponds to the
result of Matarrese et al. (2000) (but with $\delta_L\simeq 1.59$ instead of $1.686$
as explained above), agrees reasonably well with simulations.
Using the fitting function (\ref{fitfsig}) or the simple multiplicative factor
(\ref{mult}) yields close results in this regime and also agrees with simulations.
However, using Eq.(\ref{rationM}) with the fitting function (\ref{fitfsig}) appears to
agree somewhat better with simulations, especially at low masses.
This could be expected from the fact that this procedure ensures that
the mass function is properly normalized (in contrast, the simple multiplicative
factor (\ref{mult}) is greater than unity and does not reproduce the crossing of
both mass functions for $\nu \sim 1$).

On the other hand, let us point out that, contrary to some previous works, we do not
need to introduce any ad-hoc parameter $q$ (e.g., through a change of the form
$\delta_c\rightarrow \delta_c \sqrt{q}$ as in Grossi et al. 2009) to obtain a good
match with numerical simulations. This decrease in the linear threshold with
respect to the standard value $\delta_c \simeq 1.6754$ (for $\Om=0.27$) is
actually obtained in our approach by using the exact linear threshold
$\delta_L = \cF^{-1}(200) \simeq 1.59$, as explained above.
Therefore, the advantage of this procedure is that we do not need to run new
simulations for other cosmologies to obtain a fit for such a $q$-factor, since the value
$\delta_L = \cF^{-1}(200)$ can always be computed from the spherical collapse
dynamics.

\begin{figure}[htb]
\begin{center}
\epsfxsize=8 cm \epsfysize=6 cm {\epsfbox{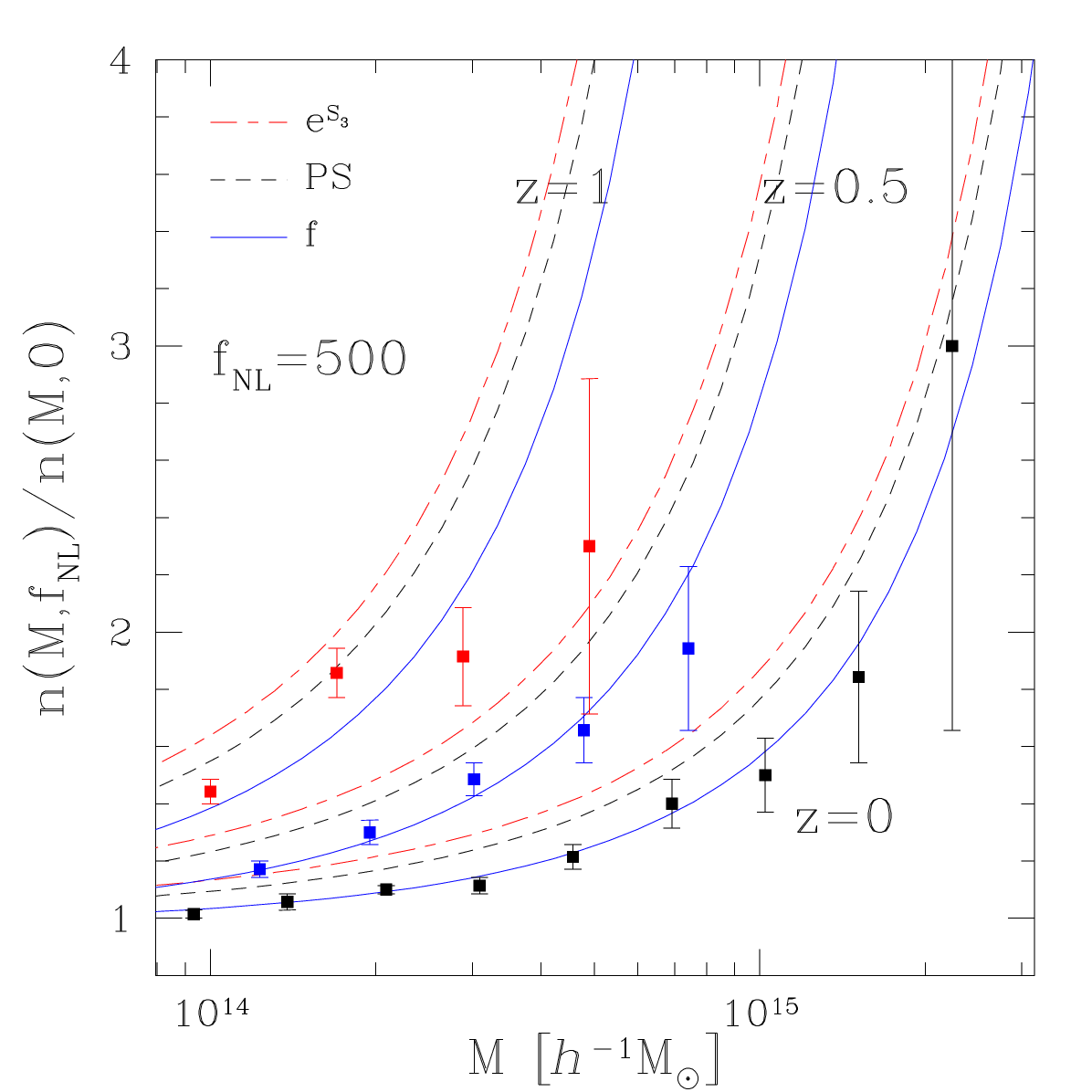}}\\
\epsfxsize=8 cm \epsfysize=6 cm {\epsfbox{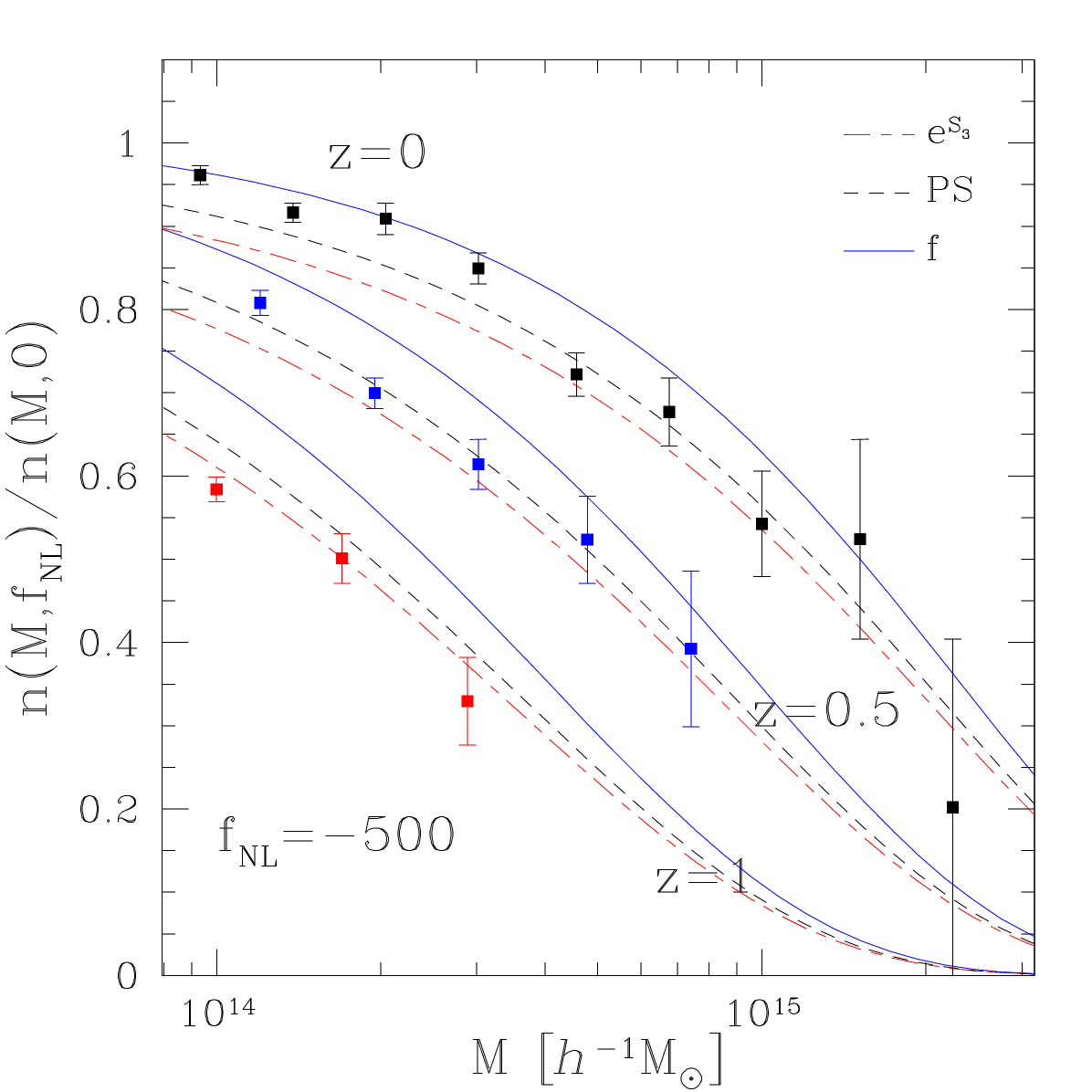}}
\end{center}
\caption{Same as Fig.~\ref{figfM_200_Grossi}, but with $\fNL=500$
(upper panel) and $\fNL=-500$ (lower panel). The data points are  the results
of the numerical simulations of Dalal et al. (2008).}
\label{figfM_Dalal}
\end{figure}

Next, we compare our results with numerical simulations from Dalal et al. (2008)
in Fig.~\ref{figfM_Dalal}. We again show the ratio $n(M,\fNL)/n(M,0)$ as a
function of mass, but for a greater primordial non-Gaussianity, $\fNL=\pm 500$.
For $\fNL=500$ (upper panel), we note as in Dalal et al. (2008) that the
prediction of (\ref{rationM}) computed with the Press-Schechter mass function
(\ref{PS3}) (i.e. the result of Matarrese et al. 2000 but with $\delta_L\simeq 1.59$)
tends to overestimate the
deviations from the Gaussian case (and the simple multiplicative factor (\ref{mult})
fares somewhat worse). However, using Eq.(\ref{rationM}) with the correct
Gaussian mass function (\ref{fitfsig}) decreases this ratio somewhat 
(as in Fig.~\ref{figfM_200_Grossi}) and provides good
agreement with the simulations.
For $\fNL=-500$ (lower panel) the match is not as good.
However, in that case the agreement
might improve at higher masses, where there are no data points but where
the predictions (\ref{mult}) or (\ref{rationM}) are asymptotically exact.
Moreover, the discrepancy between theoretical
predictions and numerical simulations has the same order as the deviation
between the different theoretical curves. Since the latter show the same
exact high-mass behavior (at the leading order given by the exponential cutoff
(\ref{PS3})), these deviations show the sensitivity of the mass functions to the
details of the theoretical prescriptions. These
power-law prefactors have not
been rigorously derived (and are expected not to be exact for all formulae used
here). Therefore, the deviation between these theoretical predictions
estimates the theoretical uncertainty for the ratio of the halo mass
functions. Then, taking this theoretical uncertainty into account, we can see
that the agreement with the numerical results is still reasonable.
For practical purposes, when one tries to derive constraints on cosmology from
observations of halo mass functions, it would be useful to consider several
theoretical prescriptions in addition to the best prediction
(\ref{rationM})-(\ref{fitfsig}),
as in Figs.~\ref{figfM_200_Grossi} and \ref{figfM_Dalal},
so as to take the theoretical uncertainty into account in the analysis.

\section{Bias of dark matter halos}
\label{Bias-of-halos}

\subsection{Two-cell saddle point and real-space bias}
\label{saddle-point2}

We now consider the bias of dark matter halos, or more precisely their two-point
correlation function. As in Valageas (2009b), following Kaiser (1984), we
identify rare massive halos with positive density fluctuations in the linear density
field. Thus, we first consider the bivariate probability distribution,
$\cP_L(\delta_{L1},\delta_{L2})$, of the linear density contrasts $\delta_{L1}$
and $\delta_{L2}$ within two spheres $V_1$ and $V_2$, of radii $q_1$ and
$q_2$, and separated by the distance
$s$. Note that we distinguish the Lagrangian distance $s$ between the halos
measured in the linear density field from their Eulerian distance $x$
measured in the nonlinear density field. Indeed, since halos have moved through
their mutual gravitational attraction, these two distances are usually different.
Proceeding as in section~\ref{saddle-point} to obtain the rare-event tails,
\beq
\cP_L(\delta_{L1},\delta_{L2}) \sim 
\max_{\{\chi[\bq]{\displaystyle |}
\delta_{Lq1}=\delta_{L1}, \delta_{Lq2}=\delta_{L2}\}} 
e^{-\frac{1}{2} \chi.C_L^{-1}.\chi} ,
\label{rareP12}
\eeq
we are led to introduce the action
\beqa
\cS[\chi,\lambda_1,\lambda_2] & = & \lambda_1 \, (\delta_{L1}- \delta_{Lq1}[\chi]) + 
\lambda_2 \, (\delta_{L2}- \delta_{Lq2}[\chi]) \nonumber \\
&& + \frac{1}{2} \, \chi.C_L^{-1}.\chi ,
\label{S12def}
\eeqa
which now involves the two Lagrange multipliers $\lambda_1$, and $\lambda_2$.
We again obtain the minimizer of the Gaussian weight
$\Gam_{12}=(\chi.C_L^{-1}.\chi)/2$
of Eq.(\ref{rareP12}) by differentiating the action $\cS$ with respect to
$\chi$, $\lambda_1$ and $\lambda_2$, and we solve these equations as a
perturbative series over $\fNL$. At order zero, we recover the Gaussian terms
\beqa
\chi^{(0)}(\bq) \!\! & = & \!\! \lambda_1^{(0)} \!\! \int_{V_1} \!\! 
\frac{\dd\bq_1}{V_1} \, C_L(\bq,\bq_1) + \lambda_2^{(0)} \!\! 
\int_{V_2} \!\! \frac{\dd\bq_2}{V_2} \, C_L(\bq,\bq_2)
\label{chi0_12} \\
\lambda_1^{(0)} & = & \frac{\sigma_2^2\,\delta_{L1}-\sigma_{12}^2\,\delta_{L2}}
{\sigma_1^2\,\sigma_2^2-\sigma_{12}^4} ,
\label{lambda0_1} \\
\lambda_2^{(0)} & = & \frac{\sigma_1^2\,\delta_{L2}-\sigma_{12}^2\,\delta_{L1}}
{\sigma_1^2\,\sigma_2^2-\sigma_{12}^4} ,
\label{lambda0_2}
\eeqa
where we note $\sigma_i^2=\sigma^2_{q_i,q_i}(0)$ and
$\sigma_{12}^2=\sigma^2_{q_1,q_2}(s)$ from Eq.(\ref{sigq1q2}).
At first order we obtain
\beqa
\chi^{(1)}(\bq) & = & \lambda_1^{(1)} \int_{V_1}\frac{\dd\bq_1}{V_1} \, 
C_L(\bq,\bq_1) + \lambda_2^{(1)} \int_{V_2}\frac{\dd\bq_2}{V_2} \, 
C_L(\bq,\bq_2) \nonumber \\
&& \hspace{-1.5cm} + 2 \lambda_1^{(0)} \!\! \int_{V_1} \!\! \frac{\dd\bq_1}{V_1} 
\! \int \!\! \dd\bq_1'\dd\bq_2' \, \fNLd(\bq_1;\bq_1',\bq_2') \, C_L(\bq,\bq_1')
\, \chi^{(0)}(\bq_2')  \nonumber \\
&& \hspace{-1.5cm} + 2 \lambda_2^{(0)} \!\! \int_{V_2} \!\! \frac{\dd\bq_2}{V_2} 
\! \int \!\! \dd\bq_1'\dd\bq_2' \, \fNLd(\bq_2;\bq_1',\bq_2') \, C_L(\bq,\bq_1')
\, \chi^{(0)}(\bq_2')  \nonumber \\
\label{chi1_12}
\eeqa
\beqa
\lambda_1^{(1)} (\sigma_1^2\sigma_2^2-\sigma_{12}^4) & = & \lambda_1^{(0)2} 
\left[ \sigma_{12}^2 (2f_{1;12}+f_{2;11}) - 3\sigma_2^2 f_{1;11} \right] 
\nonumber \\
&& \hspace{-2cm} + 2 \lambda_1^{(0)} \lambda_2^{(0)} 
\left[ \sigma_{12}^2 (2f_{2;12}+f_{1;22}) - \sigma_2^2 (2f_{1;12}+f_{2;11})
\right] \nonumber \\
&& \hspace{-2cm} + \lambda_2^{(0)2} \left[ 3\sigma_{12}^2 f_{2;22} 
- \sigma_2^2 (2f_{2;12}+f_{1;22}) \right] ,
\label{lambda1_1}
\eeqa
and
\beqa
\lambda_2^{(1)} (\sigma_1^2\sigma_2^2-\sigma_{12}^4) & = & \lambda_1^{(0)2} 
\left[ 3\sigma_{12}^2 f_{1;11} - \sigma_1^2 (2f_{1;12}+f_{2;11}) \right]
\nonumber \\
&& \hspace{-2cm} + 2 \lambda_1^{(0)} \lambda_2^{(0)} 
\left[ \sigma_{12}^2 (2f_{1;12}+f_{2;11}) - \sigma_1^2 (2f_{2;12}+f_{1;22})
\right] \nonumber \\
&& \hspace{-2cm} + \lambda_2^{(0)2} \left[ \sigma_{12}^2 (2f_{2;12}+f_{1;22}) 
- 3\sigma_1^2 f_{2;22} \right] ,
\label{lambda1_2}
\eeqa
where we note for instance $f_{1;22}=f_{a;bb}(s)$ and $f_{2;12}=f_{b;ab}(s)$,
with $V_a=V_1$, $V_b=V_2$, from Eqs.(\ref{fabb})-(\ref{fbab}).
Next, the Gaussian weight $\Gam_{12}=(\chi.C_L^{-1}.\chi)/2$ reads at order zero
\beq
\Gam_{12}^{(0)} = \frac{\sigma_2^2\,\delta_{L1}^2+\sigma_1^2\,\delta_{L2}^2 
- 2 \sigma_{12}^2 \,\delta_{L1}\delta_{L2}}
{2\,(\sigma_1^2\,\sigma_2^2-\sigma_{12}^4)} ,
\label{Gam0_12}
\eeq
and at first order,
\beqa
\Gam_{12}^{(1)} & = & \lambda_1^{(1)} \left( \sigma_1^2 \lambda_1^{(0)} 
+ \sigma_{12}^2 \lambda_2^{(0)} \right) + \lambda_2^{(1)} \left( 
\sigma_{12}^2 \lambda_1^{(0)} + \sigma_2^2 \lambda_2^{(0)} \right) 
\nonumber \\
&& + 2 \lambda_1^{(0)3} f_{1;11} + 2 \lambda_1^{(0)2}
\lambda_2^{(0)} (2 f_{1;12}+f_{2;11}) \nonumber \\
&& + 2 \lambda_1^{(0)} \lambda_2^{(0)2} 
(2 f_{2;12}+f_{1;22}) + 2 \lambda_2^{(0)3} f_{2;22} .
\label{Gam1_12}
\eeqa
Therefore, in the rare-event limit the tail of the bivariate distribution
(\ref{rareP12}) reads as
\beq
\cP_L(\delta_{L1},\delta_{L2}) \sim \cP_L(\delta_{L1}) \, \cP_L(\delta_{L2})
\, e^{-\Delta \Gam} ,
\label{P12P1P2}
\eeq
with 
\beq
\Delta\Gam = \Gam_{12} - \Gam_1 - \Gam_2 ,
\label{DGamdef}
\eeq
where $\Gam_{12}$ is given by Eqs.(\ref{Gam0_12})-(\ref{Gam1_12}) and
$\Gam_1$ and $\Gam_2$ by Eqs.(\ref{Gam0})-(\ref{Gam1}) for each sphere
$V_1$ and $V_2$.
Next, following Valageas (2009b), we write the halo two-point correlation as
\beq
1+\xi_{M_1,M_2}(x) = (1+\delta_{LM}(s)) \, e^{-\Delta \Gam} ,
\label{xiM1M2}
\eeq
where the factor $(1+\delta_{LM}(s))$ models the effects associated with the
mapping from Lagrangian to Eulerian space. This is the local linear density
contrast at radius $s$ from a halo of mass $M=\max(M_1,M_2)$ (to keep the
symmetry $M_1\leftrightarrow M_2$), as given by
Eqs.(\ref{deltap0})-(\ref{deltap1}). A sufficiently accurate approximation would
be to use only the zeroth-order term (\ref{deltap0}), as shown by
Fig.~\ref{figdeltaLq}, but taking the correction (\ref{deltap1}) into account brings
no further difficulty.
Here we approximated the nonlinear density
contrast $\delta_M$ by the linear density contrast $\delta_{LM}$, since at a
large distance where $\delta_M\ll 1$ we have $\delta_M \simeq \delta_{LM}$.
Next, we must express the Lagrangian separation $s$ in terms of the Eulerian
distance $x$. Following Valageas (2009b), at the lowest order where we consider each
halo as a test particle that falls into the potential well built by the other
halo, we obtain $s$ as the solution of the implicit equation
\beq
x= s \left( 1 - \frac{\delta_{LM_1}(s)}{3} - \frac{\delta_{LM_2}(s)}{3} \right) ,
\label{xs}
\eeq
where $\delta_{LM_i}(s)$ is the linear density contrast within radius $s$ of
the halo of mass $M_i$, given by Eqs.(\ref{delta0})-(\ref{delta1}).
At a large separation, this relation can be inverted as
\beq
s= x \left( 1 + \frac{\delta_{LM_1}(x)}{3} + \frac{\delta_{LM_2}(x)}{3} \right) ,
\label{sx}
\eeq
which provides an explicit expression for $s$.

Finally, we define the real-space halo bias as the ratio of the halo and matter
two-point correlations,
\beq
b^2_{M_1,M_2}(x) = \frac{\xi_{M_1,M_2}(x)}{\xi(x)} .
\label{biasdef}
\eeq
Since at a large distance the matter correlation is within the linear regime,
$\xi(x) \simeq \sigma^2_{0,0}(x)$, we also write in this limit
\beq
b^2_{M_1,M_2}(x) \simeq \frac{\xi_{M_1,M_2}(x)}{\sigma^2_{0,0}(x)} ,
\label{bias_sig}
\eeq
which fully determines the halo bias from Eq.(\ref{xiM1M2}).

For equal-mass halos of radius $q$, defined by the same threshold $\delta_L$,
the two Lagrange multipliers are equal, $\lambda_1=\lambda_2=\lambda$, and
Eqs.(\ref{lambda0_1})-(\ref{lambda0_2}) and (\ref{lambda1_1})-(\ref{lambda1_2})
simplify as
\beq
\lambda^{(0)} = \frac{\delta_L}{\sigma^2+\sigma_{12}^2} ,
\label{lambda0_12}
\eeq
and
\beq
\lambda^{(1)} = \frac{-3\,\delta_L^2}{(\sigma^2+\sigma_{12}^2)^3} \,
(f_{2;11}+2f_{1;12}+f_{1;11}) ,
\label{lambda1_12}
\eeq
where we note $\sigma^2=\sigma_q^2$. This yields, for the Gaussian weight
$\Gam_{12}$,
\beq
\Gam_{12}^{(0)} = \frac{\delta_L^2}{\sigma^2+\sigma_{12}^2} ,
\label{Gam0_12_q}
\eeq
and
\beq
\Gam_{12}^{(1)} = \frac{-2\,\delta_L^3}{(\sigma^2+\sigma_{12}^2)^3} \,
(f_{2;11}+2f_{1;12}+f_{1;11}) .
\label{Gam1_12_q}
\eeq
Then, the difference $\Delta\Gam$ defined in Eq.(\ref{DGamdef}), which measures
the correlation between rare events in the linear density field, writes as
\beq
\Delta\Gam^{(0)} = \frac{-\sigma^2_{12}\,\delta_L^2}
{\sigma^2\,(\sigma^2+\sigma_{12}^2)} ,
\label{DGam0}
\eeq
and
\beqa
\Delta\Gam^{(1)} & \!\!=\! & \frac{-2\,\delta_L^3}
{(\sigma^2\!+\!\sigma_{12}^2)^3} \! \left[ f_{2;11} + 2f_{1;12} 
+ \frac{\sigma^6\!-\!(\sigma^2\!+\!\sigma_{12}^2)^3}
{\sigma^6} f_{1;11} \right]  . \nonumber \\
\label{DGam1}
\eeqa
We can check that $\Delta\Gam$ vanishes for $s\rightarrow\infty$, since all mixed
quantities, $\sigma^2_{12}$, $f_{1;12}$ and $f_{2;11}$, go to zero.

As stressed in Politzer \& Wise (1984), real-space formulae such as
(\ref{xiM1M2}), which are obtained in the rare event ($\delta_L/\sigma\gg 1$)
and large separation ($\sigma_{12}\ll \sigma$) limits,
do not assume that the exponent $\Delta\Gam$ is small. In fact, as shown in 
Valageas (2009b), at high redshift one can probe a regime where this exponent
is large, so that one needs to keep the nonlinear form (\ref{xiM1M2}), which
yields a nonlinear bias. As seen from Eq.(\ref{DGam0}), this regime corresponds
to very massive halos, $\delta_L/\sigma\rightarrow\infty$, at fixed (low)
ratio $\sigma_{12}/\sigma$.
Nevertheless, in the regime where $\Delta\Gam$ is small (i.e. in the large
separation limit $x\rightarrow \infty$), which covers the cases
of interest encountered at low redshift, we can expand the exponential in
Eq.(\ref{xiM1M2}). Then, at the lowest order over the terms that vanish in the
large separation limit, we obtain
\beq
x\rightarrow \infty : \;\;\; \xi_M(x) \simeq \delta_{LM}(s) - \Delta\Gam 
\eeq
for equal-mass halos, whence
\beqa
\xi_M(x) & \!\! \simeq \! & \frac{\delta_L}{\sigma_q^2} \, \sigma^2_{q,0}(s) +
\frac{\delta_L^2}{\sigma_q^4} \,\! \left[ \! f_{0;qq}(s) \!+\! 2 f_{q;0q}(s) 
\!-\! 3 \frac{f_{q;qq}}{\sigma_q^2} \sigma^2_{q,0}(s) \right] \nonumber \\
&& \hspace{-1.2cm} + \frac{\delta_L^2}{\sigma_q^4} \sigma^2_{q,q}(s) 
+ 2 \frac{\delta_L^3}{\sigma_q^6} \left[ f_{2;11}(s) \!+\! 2 f_{1;12}(s) 
\!-\! 3 \frac{f_{1;11}}{\sigma_q^2} \sigma^2_{q,q}(s) \right] .
\label{xiMlin}
\eeqa
Hereafter we call the bias $b^2_M(x)$ obtained from Eq.(\ref{xiMlin}) and
Eq.(\ref{bias_sig}) the ``linearized'' bias.

Following the approach of Kaiser (1984), Matarrese \& Verde
(2008) also computed the effect of local-type non-Gaussianity (\ref{fNLdef}) on
the halo two-point correlation. Drawing on earlier work by Matarrese et al. (1986),
who computed the $n$-point halo correlations by expanding the relevant
path-integrals and next resumming the series within a large-distance and
rare-event approximation, they obtained expressions of the form (\ref{xiM1M2})
without the prefactor $(1+\delta_{LM})$ and (\ref{xiMlin}) without the terms in
the first line, which arise from this prefactor, and the last term $f_{1;11}$ in the
second bracket. This term arises from the factor $\sigma_{12}^2$ in the
denominator of expression (\ref{Gam1_12_q}), associated with the effect of
primordial non-Gaussianity on the one-point distribution (\ref{Gam1}).
It can be seen as a renormalization of the Gaussian term
$(\delta_L^2/\sigma_q^4) \sigma^2_{q,q}(s)$, since it shows the same scale
dependence through the function $\sigma^2_{q,q}(s)$.
The presence of such a term has already been noticed in Slosar et al. (2008)
in Fourier space, and Desjacques et al. (2009) points out that it needs to be
included to obtain good agreement with numerical simulations. This will give
rise to the last term in Eq.(\ref{PkM}) and the second term in Eqs.(\ref{bMk})
and (\ref{bMk1}) below. On the other hand, the terms in the first line of Eq.(\ref{xiMlin}),
associated with the prefactor $(1+\delta_{LM})$ in Eq.(\ref{xiM1M2}), stem from
the mapping from Lagrangian to Eulerian space, and the first bracket in
Eq.(\ref{xiMlin}) expresses the (weak) effect of primordial non-Gaussianity
on this mapping. Another difference between Eqs.(\ref{xiM1M2}) and (\ref{xiMlin})
and previous works is that (within some approximation) we pay attention to
the difference between Lagrangian and Eulerian distances $s$ and $x$.
This can play a non-negligible role as seen in Fig.~\ref{figbiasM_z0_r50} below.

\begin{figure}[htb]
\begin{center}
\epsfxsize=8.5 cm \epsfysize=6.3 cm {\epsfbox{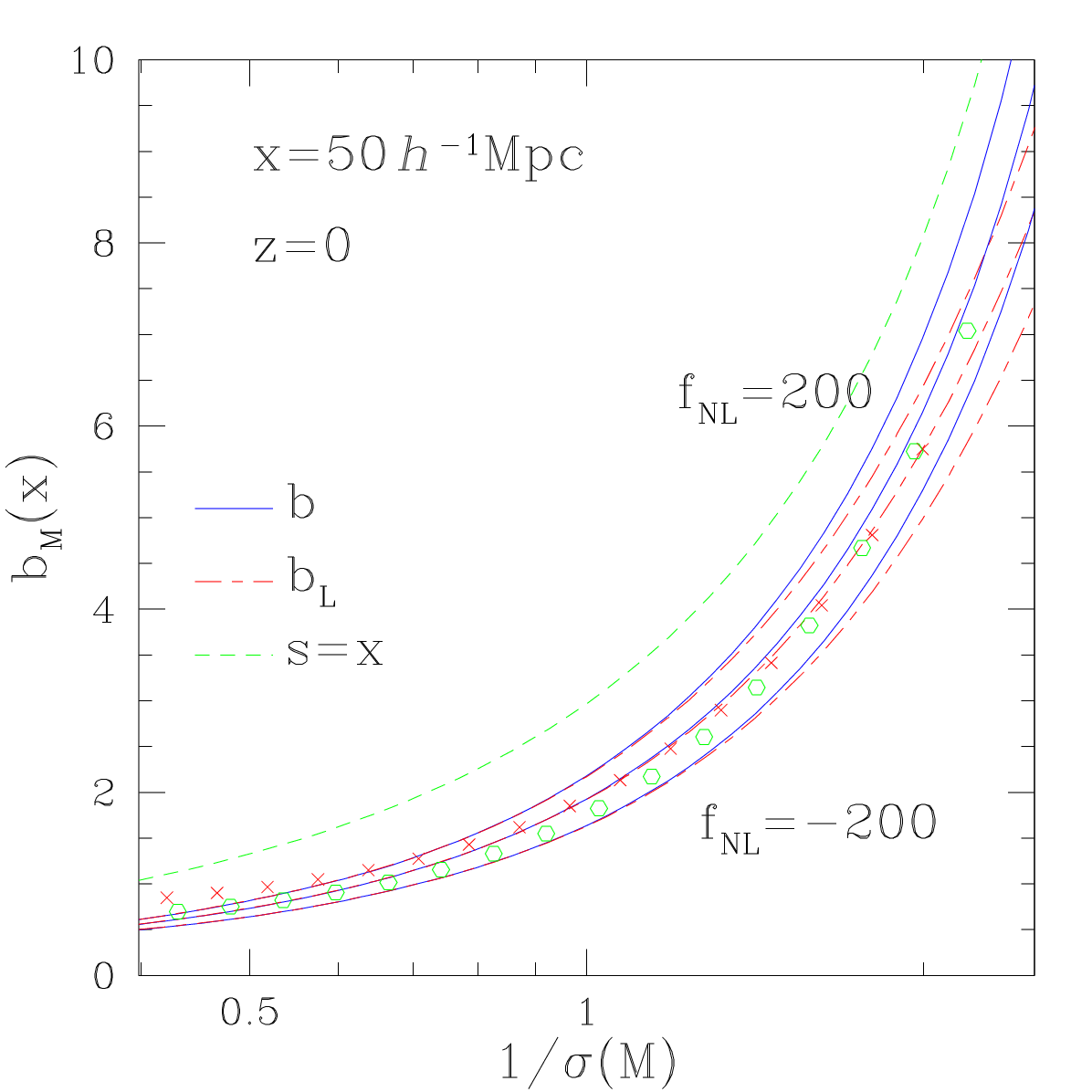}}
\end{center}
\caption{The halo bias $b_M(x)$, as a function of $\sigma(M)$, at fixed redshift
$z=0$ and distance $x=50 h^{-1}$Mpc. The solid lines ``$b$'' are the
nonlinear theoretical prediction of Eqs.(\ref{xiM1M2}), (\ref{xs}) and (\ref{bias_sig}),
for $\fNL=\pm 200$ and $\fNL=0$ (i.e. Gaussian case, intermediate line), while
the dot-dashed lines ''$b_L$'' are the linearized bias of Eq.(\ref{xiMlin}).
The upper dashed line ``$s=x$'' shows the result obtained in the Gaussian case by
setting $s=x$ in Eq.(\ref{xiM1M2}).
The points are the fits to Gaussian numerical simulations,
from Sheth, Mo \& Tormen (2001) (crosses) and Pillepich (2010) (circles).}
\label{figbiasM_z0_r50}
\end{figure}

\begin{figure}[htb]
\begin{center}
\epsfxsize=8.5 cm \epsfysize=6.3 cm {\epsfbox{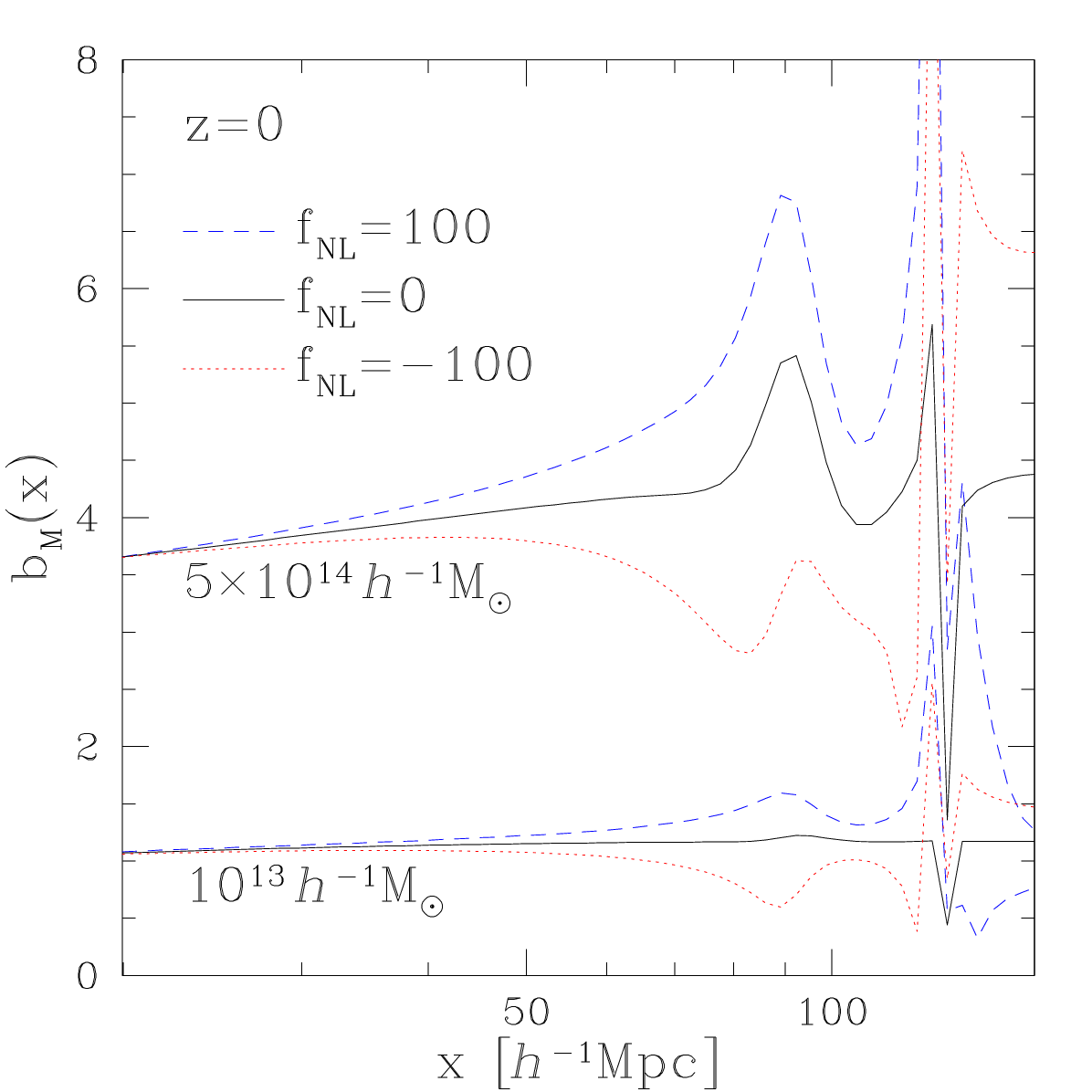}}
\epsfxsize=8.5 cm \epsfysize=6.3 cm {\epsfbox{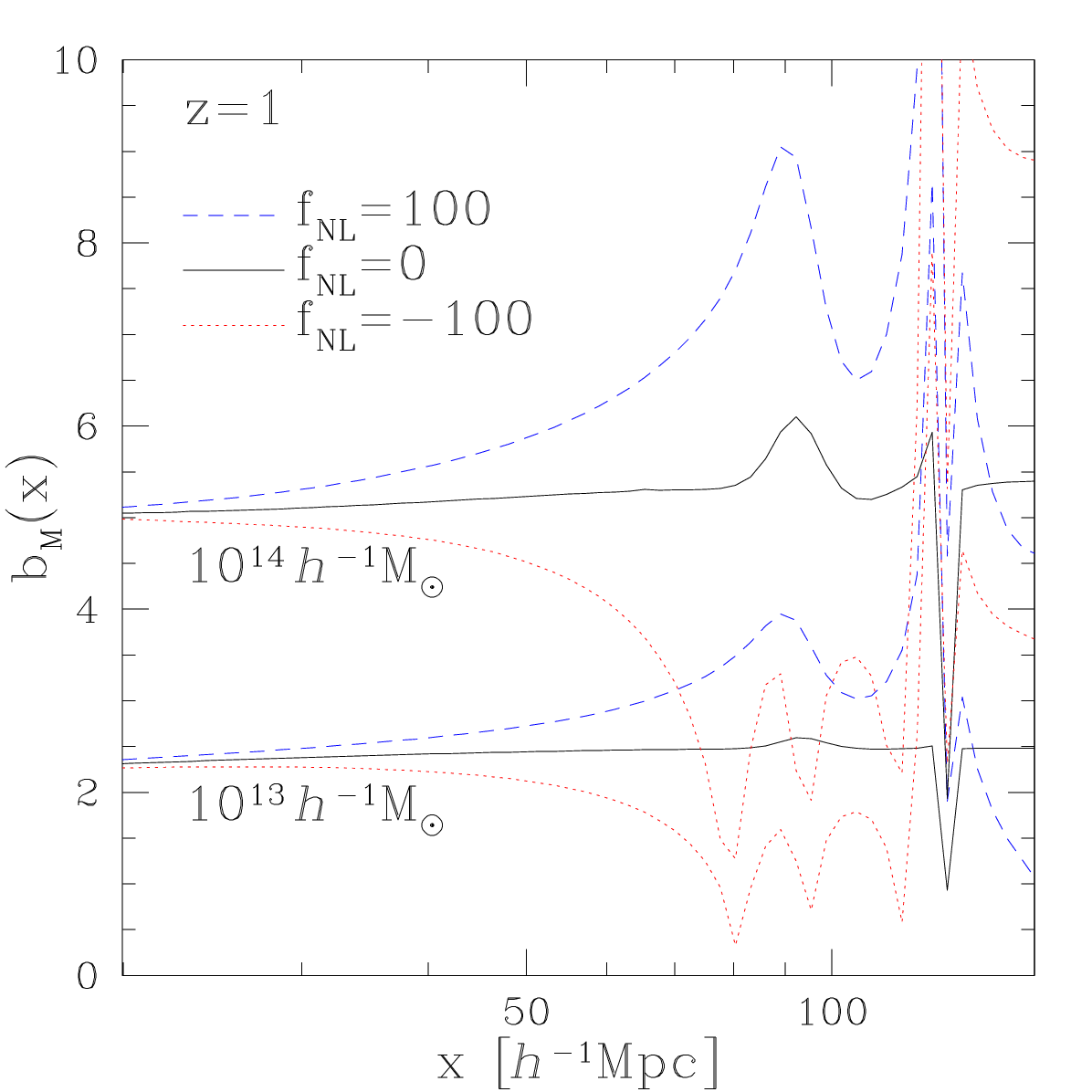}}
\end{center}
\caption{The halo bias $b_M(x)$ as a function of distance $x$, at redshifts
$z=0$ (upper panel) and $z=1$ (lower panel) for several masses.
We show the cases $\fNL= 0$ (solid lines),
$\fNL=100$ (dashed lines), and $\fNL=-100$ (dotted lines). The divergences at
$x \sim 120 h^{-1}$Mpc come from the halo and matter correlations
not changing sign at the same distance.}
\label{figbiasr}
\end{figure}

\begin{figure}[htb]
\begin{center}
\epsfxsize=8.5 cm \epsfysize=6.3 cm {\epsfbox{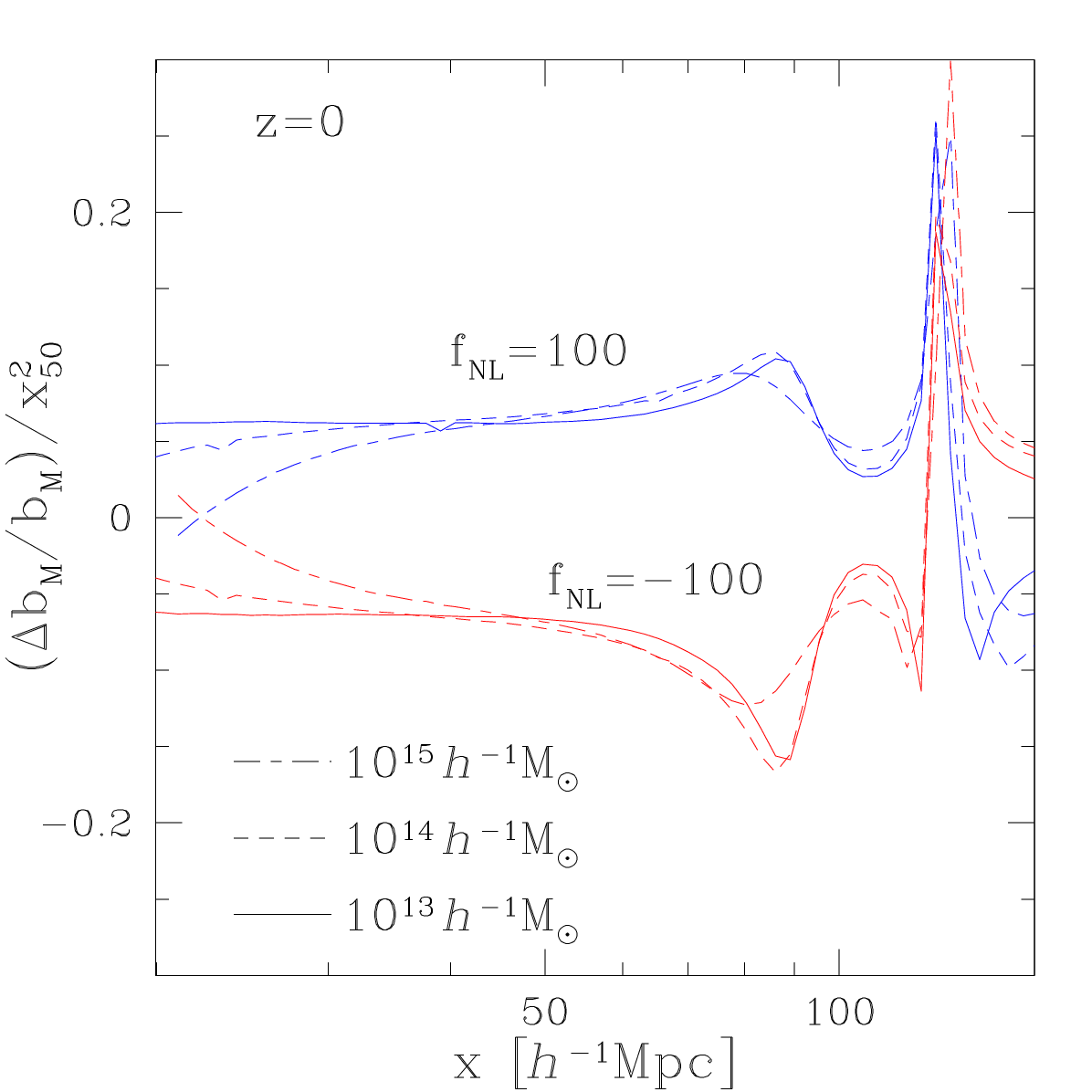}}
\epsfxsize=8.5 cm \epsfysize=6.3 cm {\epsfbox{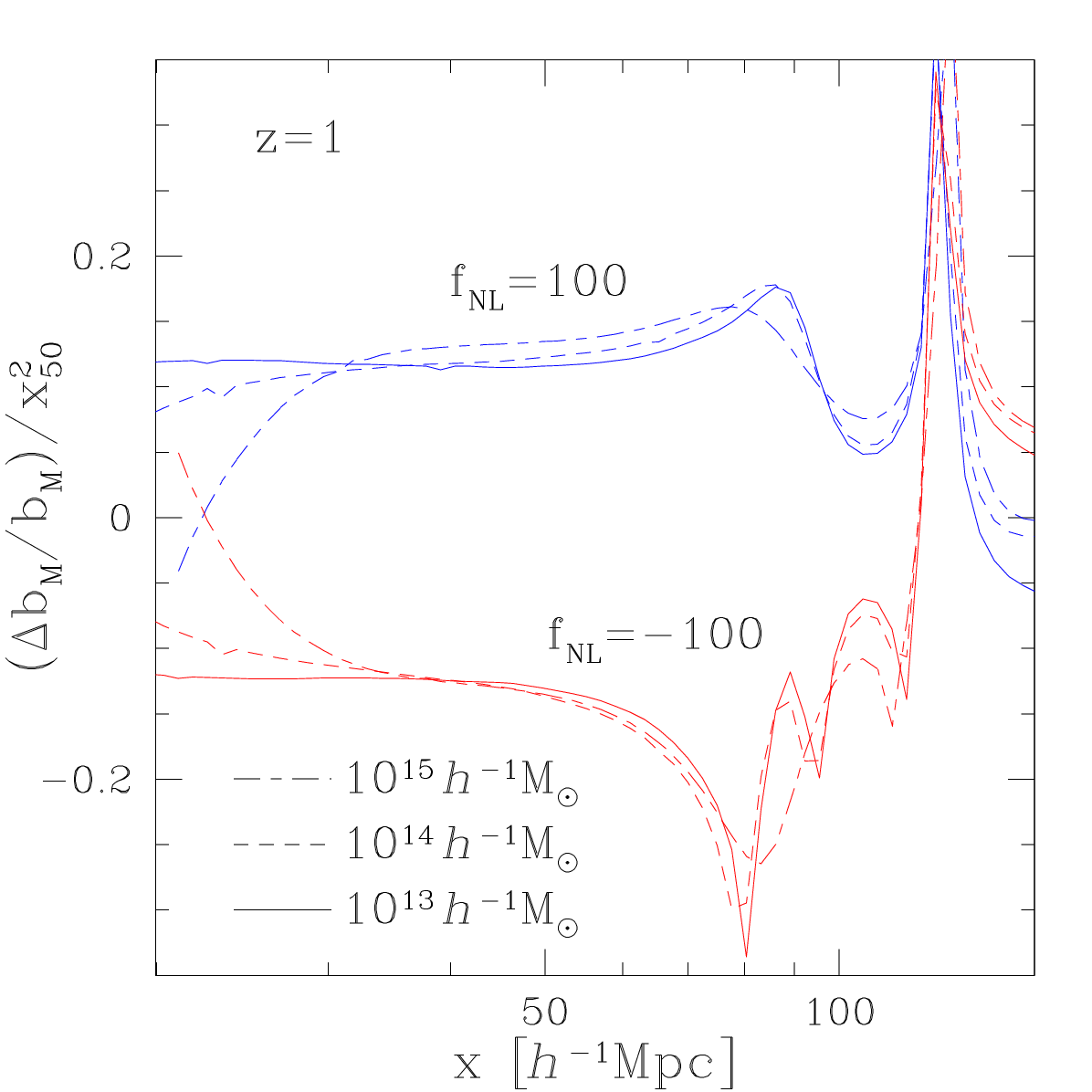}}
\end{center}
\caption{The real-space ratio $[\Delta b_M(x,\fNL)/b_M(x,0)]/x_{50}^2$ of the
correction $\Delta b_M(x,\fNL)=b_M(x,\fNL)-b_M(x,0)$ to the Gaussian bias
$b_M(x,0)$, divided by the factor $x_{50}^2$ with $x_{50}=x/(50 h^{-1} {\rm Mpc})$,
as a function of distance $x$. We show the cases $\fNL=100$ (upper lines) and
$\fNL=-100$ (lower lines) for several masses at redshifts $z=0$ (upper panel) and
$z=1$ (lower panel), from Eq.(\ref{xiM1M2}).}
\label{figDeltabiasr}
\end{figure}

We show in Fig.~\ref{figbiasM_z0_r50} the halo bias $b_M(x)$ as a function
of $\sigma(M)$ at fixed distance $x=50 h^{-1} $ Mpc and redshift $z=0$,
for $\fNL=\pm 200$ and $\fNL=0$.
We display both the nonlinear result of Eq.(\ref{xiM1M2})
and the linear result of Eq.(\ref{xiMlin}). In addition, for the
Gaussian case we also display the bias obtained by setting $s=x$ in
Eq.(\ref{xiM1M2}).
As in Valageas (2009b), for the Gaussian case ($\fNL=0$) we obtain good
agreement with the fits to numerical simulations of Sheth, Mo \& Tormen (2001)
and Pillepich (2010). Moreover, we can see that it is important to take the displacement
of the halos into account through Eq.(\ref{xs}), as the approximation $s=x$
significantly overestimates the bias. We checked that 
using the simpler Eq.(\ref{sx}) gives a close result to the one obtained with
Eq.(\ref{xs}) and also agrees with the simulations. Thus, for practical purposes
it is sufficient to use Eq.(\ref{sx}).
We can see that for all cases shown in Fig.~\ref{figbiasM_z0_r50} the linear
bias from Eq.(\ref{xiMlin}) gives results that are very close to the fully nonlinear
expression (\ref{xiM1M2}). This justifies the use of such linearized expressions
in this regime. This will be especially useful in section~\ref{Fourier-bias}, where we
consider the bias of dark matter halos in Fourier space. Indeed, it is easier
to take the Fourier transform of Eq.(\ref{xiMlin}), which allows us to recover the
results obtained in previous works. Thus, one interest of the real-space results
(\ref{xiM1M2})-(\ref{xiMlin}) is to provide a check on whether linearized predictions
(i.e. where the halo correlation only involves the matter power spectrum at linear order)
are valid. Then, in agreement with those studies (which were mostly performed in
Fourier space and led to Eq.(\ref{Dbk_Dal}), see Dalal et al. 2008; Slosar et al. 2008),
and with Eq.(\ref{xiMlin}), we can see that the deviation from the Gaussian bias,
$b_M(x,\fNL)-b_M(x,0)$, grows linearly with $b_M(x,0)$ at large bias and has the
same sign as $\fNL$.

\begin{figure}[htb]
\begin{center}
\epsfxsize=8.5 cm \epsfysize=6.3 cm {\epsfbox{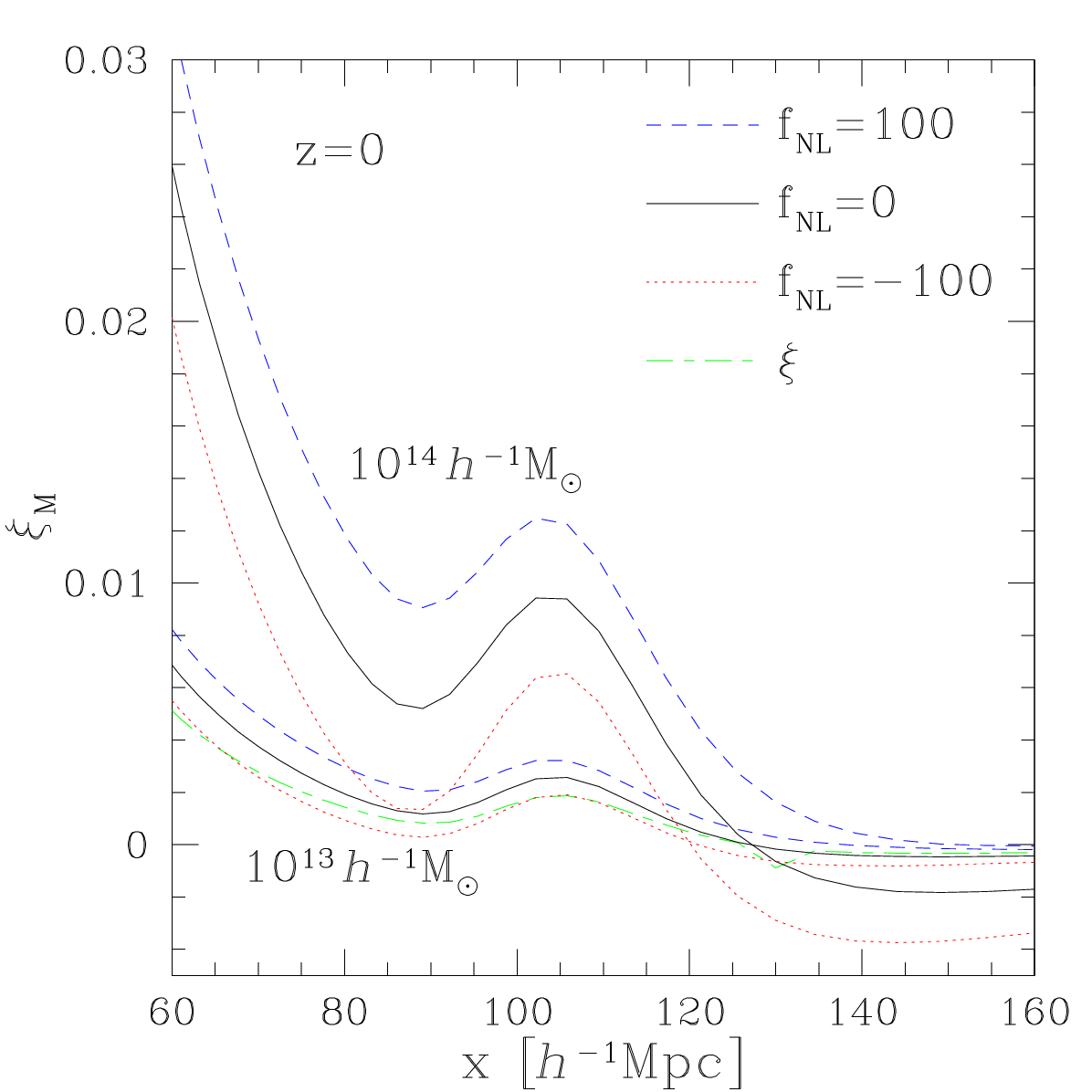}}\\
\epsfxsize=8.5 cm \epsfysize=6.3 cm {\epsfbox{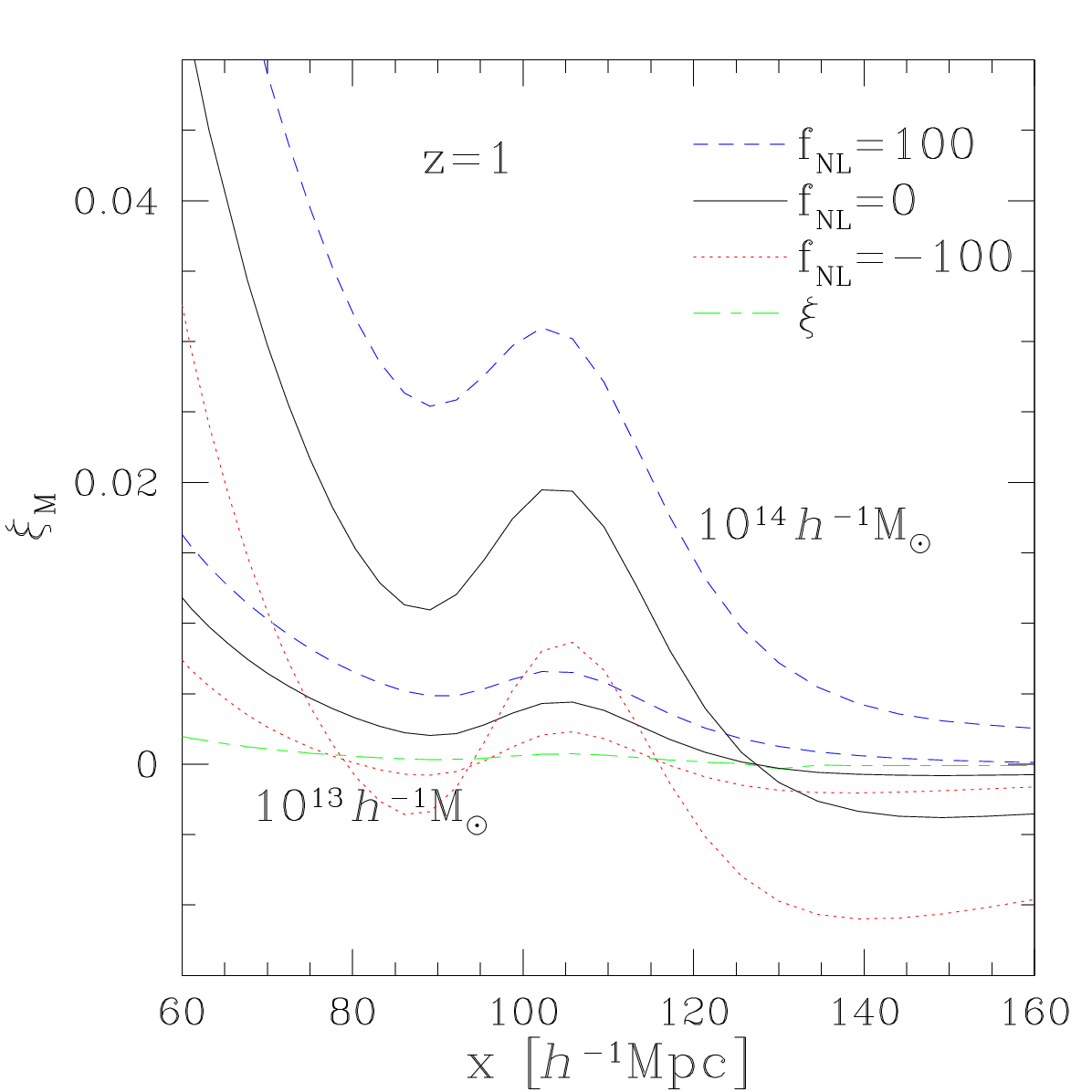}}
\end{center}
\caption{The halo ($\fNL=0, \pm 100$) and matter ($\xi$) two-point correlations
at redshifts $z=0$ (upper panel) and $z=1$ (lower panel). We show the curves
obtained for the masses $M=10^{13}$ and $10^{14} h^{-1} M_{\odot}$.}
\label{figxi}
\end{figure}

Next, we display in Fig.~\ref{figbiasr} the dependence on the distance $x$
of the bias obtained for several masses at redshifts $z=0$ and $z=1$. More precisely,
we show the ratio $\sqrt{|\xi_M(x)/\xi(x)|}$, since the halo and matter correlations
do not change sign at the same point.
We plot the cases $\fNL=\pm 100$, as well as the Gaussian case $\fNL=0$.
While the Gaussian bias is roughly constant on large scales, up to
$\sim 100 h^{-1}$Mpc (in agreement with previous studies, Mo \& White 1996;
Mo et al. 1997), the non-Gaussian bias shows a strong scale dependence,
with a deviation from the Gaussian bias that roughly grows as $x^2$ up to
$\sim 100 h^{-1}$Mpc.
This agrees with the $k^{-2}$ behavior observed in Fourier space, see
Eq.(\ref{Dbk_Dal}) above (Dalal et al. 2008; Slosar et al. 2008) and Eq.(\ref{bMk1})
below.

To see the scaling of the real-space correction
$\Delta b_M(x,\fNL)= b_M(x,\fNL)-b_M(x,0)$ to the Gaussian bias more clearly,
we show in Fig.~\ref{figDeltabiasr} the ratio
$[\Delta b_M(x,\fNL)/b_M(x,0)]/x_{50}^2$, with $x_{50}=x/(50 h^{-1}{\rm Mpc})$.
We display the results obtained from Eq.(\ref{xiM1M2}) for several masses at
$z=0$ (upper panel) and $z=1$ (lower panel) for $\fNL=\pm 100$.
We can see that over the range $30<x<90 h^{-1}$Mpc all curves roughly collapse
onto one another. This means that the real-space correction roughly scales as
$\Delta b_M(x) \propto \fNL b_M(x,0) x^2$ over this range,
which roughly agrees with the Fourier-space scaling (\ref{Dbk_Dal}) (here we
neglected any constant offset, such as the factor $-1$ in Eq.(\ref{Dbk_Dal}), see
the discussion of Eq.(\ref{Delta0bk}) below). It appears that our predictions
scale more closely as $x^2$, as shown in Fig.~\ref{figDeltabiasr}, than as
$\alpha(1/x)^{-1}$, which would be suggested by Eqs.(\ref{Dbk_Dal}), (\ref{bMk1})
(at these scales the transfer function already deviates from unity).
This agrees with the behavior observed in Fourier space in Fig.~\ref{figDeltabiask}
below.
The masses shown in Fig.~\ref{figDeltabiasr} span the range
$1.2<b_M(x,0)<5.9$ at $x=50 h^{-1}$Mpc and $z=0$, and $2.4<b_M(x,0)<15.4$
at $z=1$, so that the linear scaling with
$b_M(x,0)$ of the correction $\Delta b_M(x,\fNL)$ appears to be a good
approximation.

Below $30 h^{-1}$Mpc higher masses show steeper scale dependence
for $\Delta b_M(x,\fNL)$.
At very large distance, $x > 100 h^{-1}$Mpc, the oscillations seen in 
Figs.~\ref{figbiasr} and \ref{figDeltabiasr} are caused by the 
baryon acoustic oscillation. Indeed, the baryon oscillations seen in the halo and
matter two-point correlations are not exactly proportional, since the halo correlation
is not exactly proportional to $\sigma^2_{0,0}(x)$, even in the Gaussian case and
in the linear regime (for instance it involves the smoothing scale $q$,
see Eq.(\ref{xiMlin})). This yields the non-monotonic behavior seen in
Figs.~\ref{figbiasr} and \ref{figDeltabiasr} around $100 h^{-1}$Mpc.
For the same reason, the halo
and matter correlations do not exactly vanish at the same distance, which gives
rise to the divergent spike at $\sim 120 h^{-1}$ Mpc.
These features simply mean that it is no longer useful to work with the bias $b_M$
on these scales, which only makes sense if the halo and matter correlations are
roughly proportional. In this range, where the correlations show some oscillations
and change sign, it is no longer a good approximation to write the halo correlation
in terms of the matter correlation multiplied by some slowly varying bias factor.
Then, one instead needs to directly study the halo and matter correlations themselves.

Thus, we compare in Fig.~\ref{figxi} the halo and matter two-point correlations.
We focus on large scales to see how the baryon acoustic oscillation is modified
when one uses massive halos as a tracer of the initial matter power spectrum.
In agreement with previous works (Desjacques 2008), we can see that the
oscillation is strongly amplified for massive halos that have a strong bias.
This amplification still holds for significant primordial non-Gaussianity ($\fNL=\pm 100$),
although it appears to be slightly lower for positive $\fNL$.
Moreover, the peak of the oscillation shows no significant shift, so that a measure
of its position appears to be a robust ruler for constraining cosmology, independently
of the halo bias and of the primordial non-Gaussianity.
In contrast, the distance at which the two-point correlation changes
sign is not significantly modified as one goes from the matter to the halo correlation
in the Gaussian case, but it is fairly sensitive to the primordial non-Gaussianity.
In particular, a positive $\fNL$ shifts this point to a greater distance.
However, theoretical and observational error bars may be too large to use this
effect to constrain $\fNL$ in a competitive manner compared to other probes.

\subsection{Fourier-space bias}
\label{Fourier-bias}

Rather than the real-space two-point correlation, recent works have mostly studied
the effect of primordial non-Gaussianity on the halo power spectrum, where at
lowest order the Poisson equation (\ref{alphadef}) directly gives an estimate of the
form (\ref{Dbk_Dal}) for the deviation from the
bias obtained with Gaussian initial conditions (Dalal et al. 2008;
Slosar et al. 2008).

It is not convenient to take the Fourier transform of the nonlinear correlation
(\ref{xiM1M2}), but at moderate redshifts, the linearized
form (\ref{xiMlin}) provides a very good approximation. Then, if we also
make the approximation $s \simeq x$, which is valid at the lowest order,
the Fourier transform of Eq.(\ref{xiMlin}) readily gives the halo power spectrum
as
\beqa
P_M(k) & \simeq & \frac{\delta_L}{\sigma_q^2} P_L(k) \tW(kq) +
\frac{\delta_L^2}{\sigma_q^4} \biggl[ \tf_{0;qq}(k) + 2 \tf_{q;0q}(k) 
\nonumber \\
&& \hspace{-1cm} - 3 \frac{f_{q;qq}}{\sigma_q^2} P_L(k) \tW(k q) \biggl]
+ \frac{\delta_L^2}{\sigma_q^4}  P_L(k) \tW(kq)^2 \nonumber \\
&& \hspace{-1cm} 
+ 2 \frac{\delta_L^3}{\sigma_q^6} \left[ \tf_{2;11}(k) \!+\! 2 \tf_{1;12}(k) 
\!-\! 3 \frac{f_{1;11}}{\sigma_q^2} P_L(k) \tW(kq)^2 \right] ,
\label{PkM}
\eeqa
where the quantities $\tf$ are obtained from Eqs.(\ref{tfabb})-(\ref{tfbab}).
As discussed below Eq.(\ref{xiMlin}), the terms $\tf_{2;11}(k)$ and $2\tf_{1;12}(k)$
in Eq.(\ref{PkM}) have already been obtained by Matarrese \& Verde (2008),
following Kaiser (1984) by identifying massive halos with rare fluctuations
in the linear density field.
Defining the Fourier-space bias as (note that this is not the Fourier transform
of the real-space bias (\ref{biasdef}))
\beq
b_M^2(k) = \frac{P_M(k)}{P(k)} \simeq \frac{P_M(k)}{P_L(k)} ,
\label{bkdef}
\eeq
where we used $P(k) \simeq P_L(k)$ at low $k$ for the matter power spectrum,
we obtain $b_M^2(k)$ from Eq.(\ref{PkM}). We can also obtain the bias of
different-mass halos in a similar fashion, first expanding Eq.(\ref{xiM1M2})
and next taking the Fourier transform.
To consider the displacement of the halos (i.e. $x \neq s$),
we can also make the approximation
\beq
P_M(k) \rightarrow \left(\frac{x}{s}\right)^3 P_M\left(\frac{x}{s}k\right) ,
\;\; \mbox{with} \;\;  x= \frac{1}{k} ,
\label{PkMs}
\eeq
where we use the explicit expression (\ref{sx}) for $s(x)$.
This expresses that Lagrangian-space wavelengths
$\sim s$ (i.e. measured in the linear density field) correspond to smaller
Eulerian-space wavelengths $\sim x$ (i.e. measured in the nonlinear density field)
because of the displacement of massive halos, which usually have come
closer because of their mutual attraction ($x<s$).
This also follows the spirit of the Hamilton et al. (1991) ansatz, translated in Fourier
space in Peacock \& Dodds (1996).

For large negative $\fNL$ the halo power spectrum (\ref{PkM})
can become negative, because we looked for an expression
of the halo correlation, or of the halo power spectrum, at linear order over $\fNL$
as in Eq.(\ref{PkM}). However, the power spectrum $P_M(k)$ must be positive
by definition. Then, in cases where expression (\ref{PkM}) turns negative
one should consider higher-order terms over $\fNL$, which would ensure that
the power spectrum remains positive.
Nevertheless, since such high-order terms are beyond the scope of this article,
we consider below the following simple procedure that ensures that $P_M(k)$,
or the squared bias $b_M^2(k)$, remain positive.
Up to linear order over $\fNL$, the bias (\ref{bkdef}) reads as
\beqa
b_M(k,\fNL) & = & \sqrt{\frac{P_M(k,0)+\Delta P_M(k,\fNL)}{P(k)}} \nonumber \\
& \simeq & b_M(k,0) \, \left( 1 + \frac{1}{2} \frac{\Delta P_M(k,\fNL)}{P_M(k,0)} \right) ,
\label{Dbk}
\eeqa
where $\Delta P_M(k,\fNL) = P_M(k,\fNL) - P_M(k,0)$ is the deviation from the
Gaussian halo power spectrum, and we expanded the square-root. Then, we may use
the last expression (\ref{Dbk}) as the prediction of the halo bias. This amounts
to making the transformation
\beq
b_M^2(k,\fNL) \rightarrow b_M^2(k,0) \, \left[ 1 + \frac{1}{2} \left( 
\frac{b_M^2(k,\fNL)}{b_M^2(k,0)} - 1 \right) \right]^2 ,
\label{bk1}
\eeq
where the two sides only differ by higher-order terms over $\fNL$ (as
$1+\epsilon \simeq (1+\epsilon/2)^2$ up to linear order over $\epsilon$)
and the right side is always positive.

\begin{figure}[htb]
\begin{center}
\epsfxsize=4.44 cm \epsfysize=4.5 cm {\epsfbox{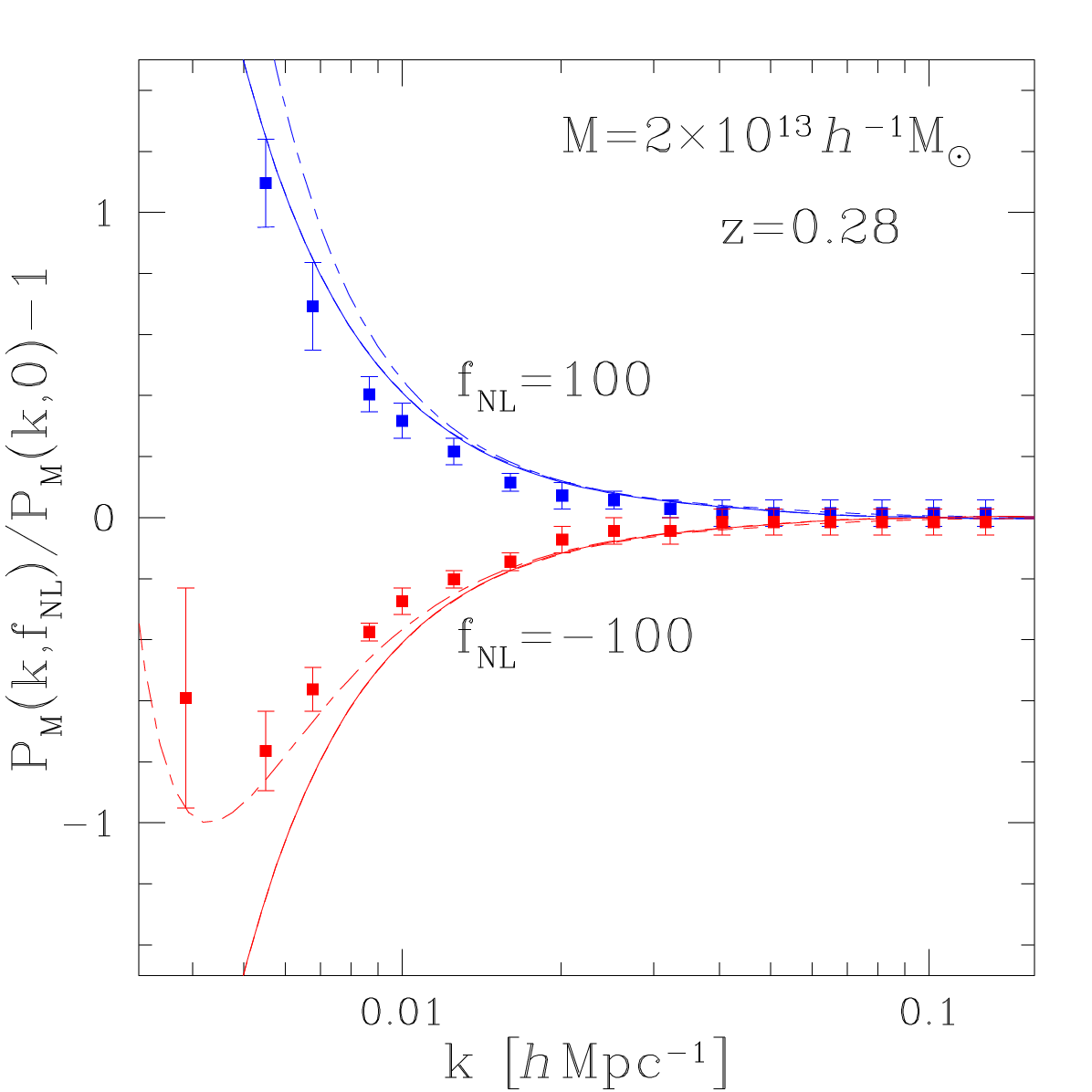}}
\epsfxsize=4.44 cm \epsfysize=4.5 cm {\epsfbox{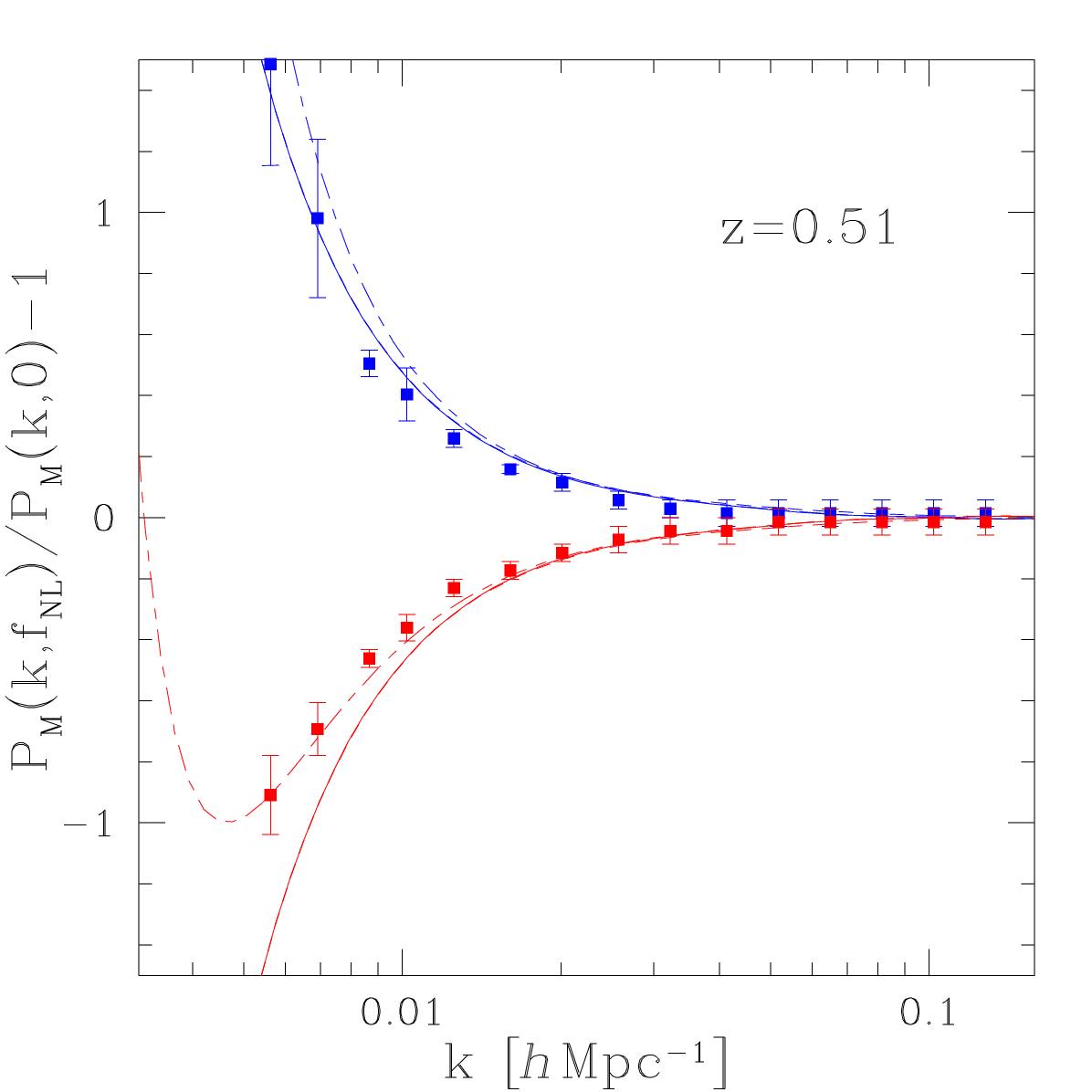}}\\
\epsfxsize=4.44 cm \epsfysize=4.5 cm {\epsfbox{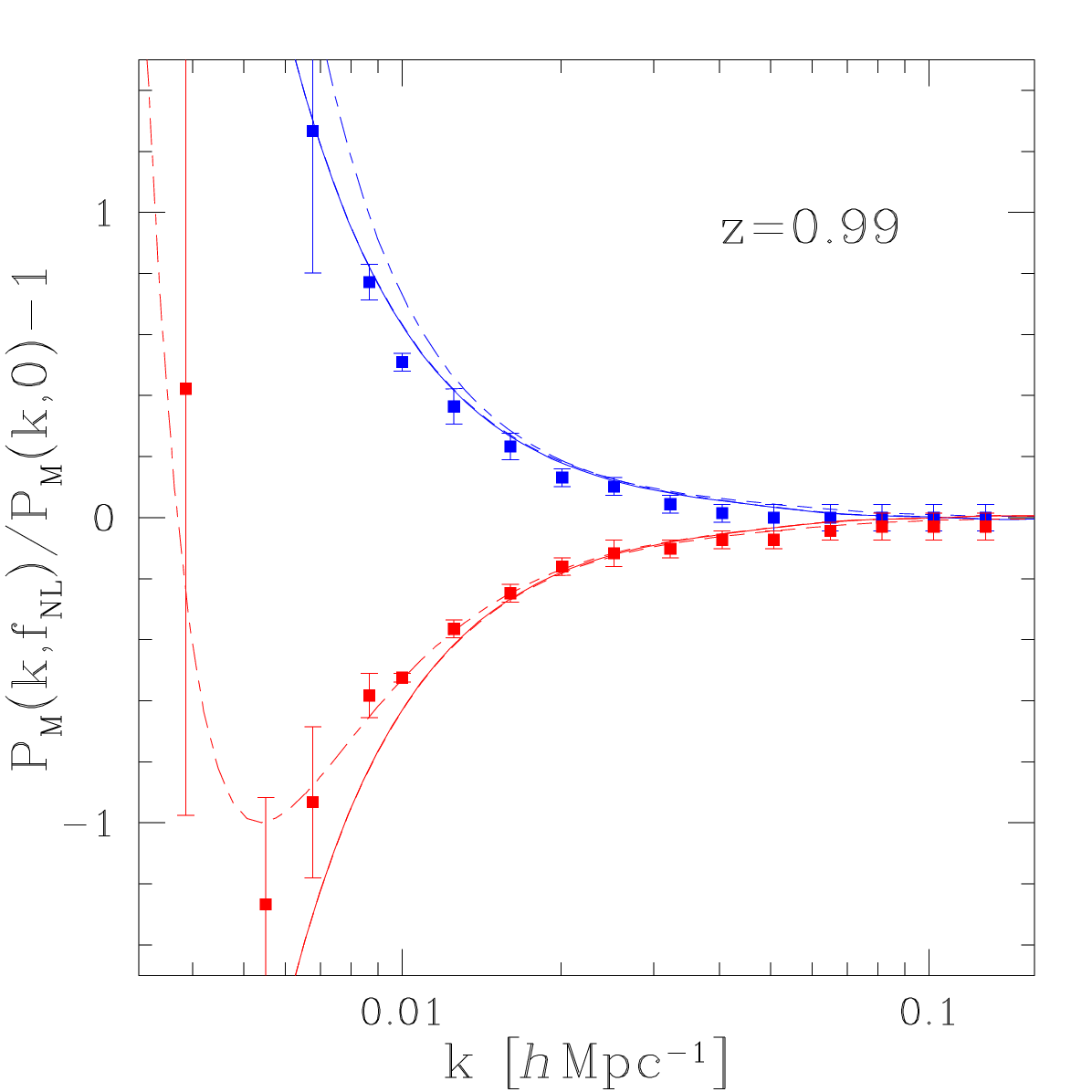}}
\epsfxsize=4.44 cm \epsfysize=4.5 cm {\epsfbox{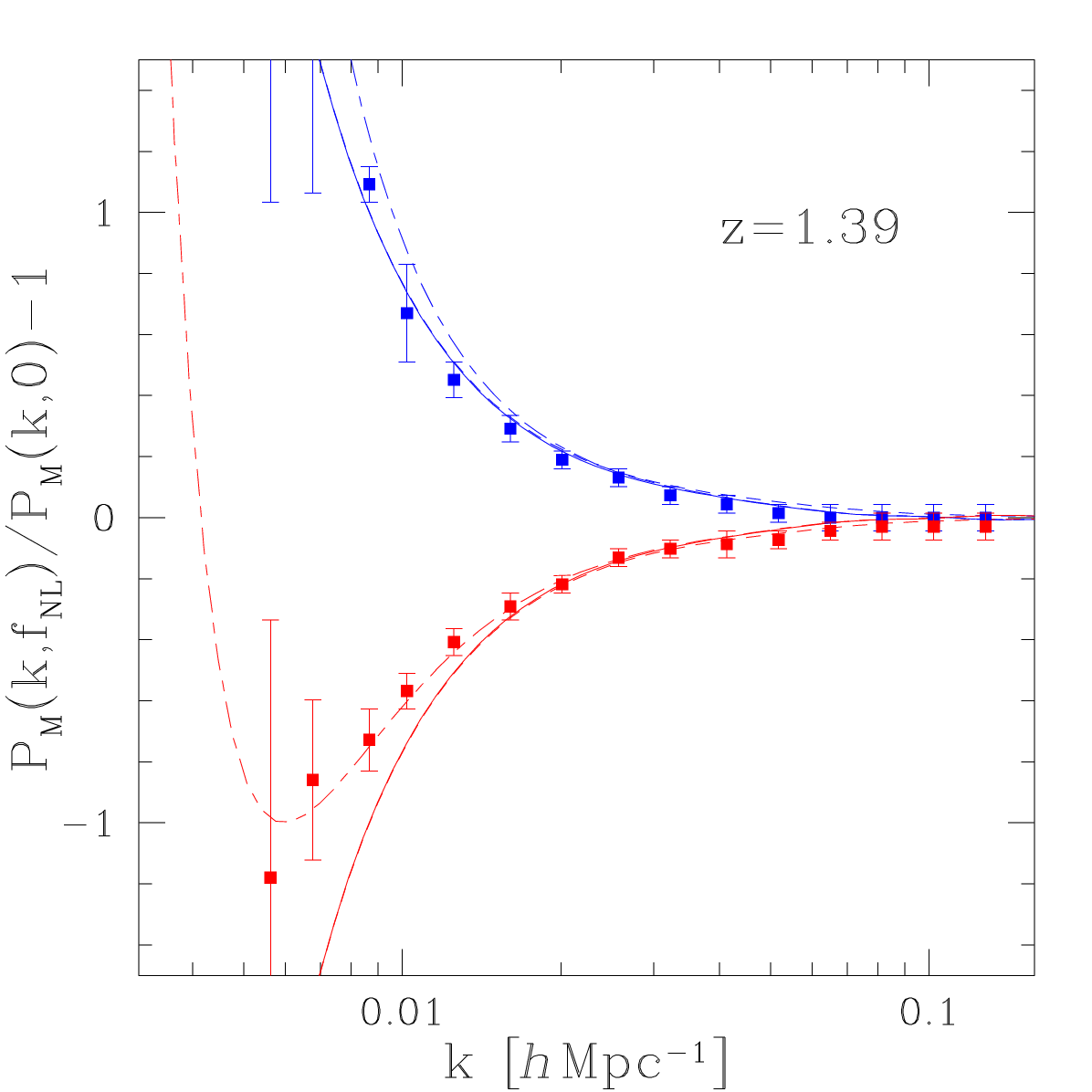}}\\
\epsfxsize=4.44 cm \epsfysize=4.5 cm {\epsfbox{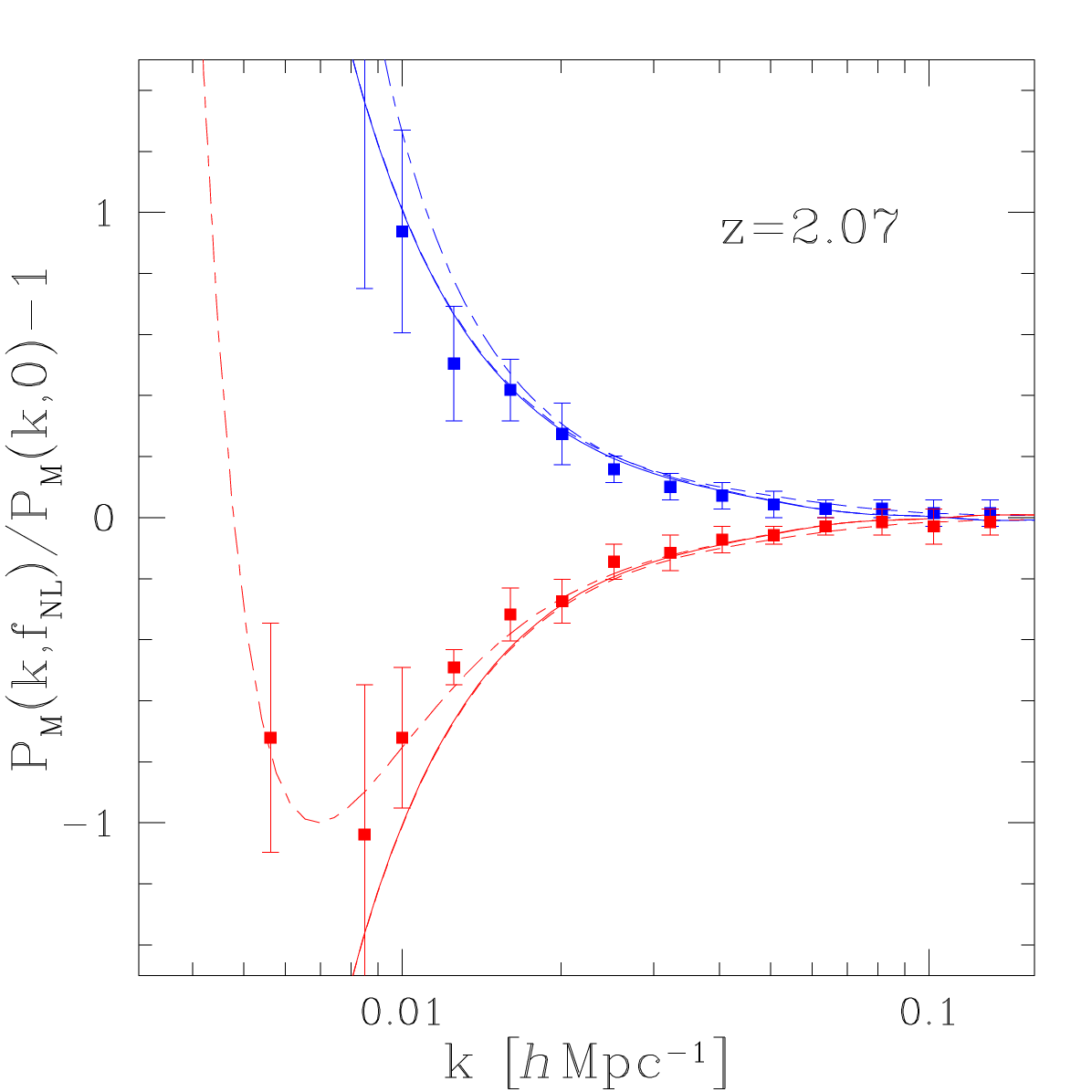}}
\end{center}
\caption{The correction to the halo power spectrum due to primordial
non-Gaussianity. We show the ratio $P_M(k,\fNL)/P_M(k,0)-1$ for $\fNL=100$
(upper curves) and $\fNL=-100$ (lower curves). The solid and dashed curves that
are almost indistinguishable are Eqs.(\ref{PkM}) and (\ref{PkMs}). The dot-dashed
curves that are above the solid curves at low $k$ correspond to Eq.(\ref{bk1})
which ensures that the halo power spectrum is always positive. The theoretical
predictions are for $M=2\times 10^{13}h^{-1} M_{\odot}$ and the data points
are the numerical simulations of Desjacques et al. (2009), for
$M>2\times 10^{13}h^{-1} M_{\odot}$.}
\label{figbiask_M2d13_Desj}
\end{figure}

\begin{figure}[htb]
\begin{center}
\epsfxsize=4.44 cm \epsfysize=4.5 cm {\epsfbox{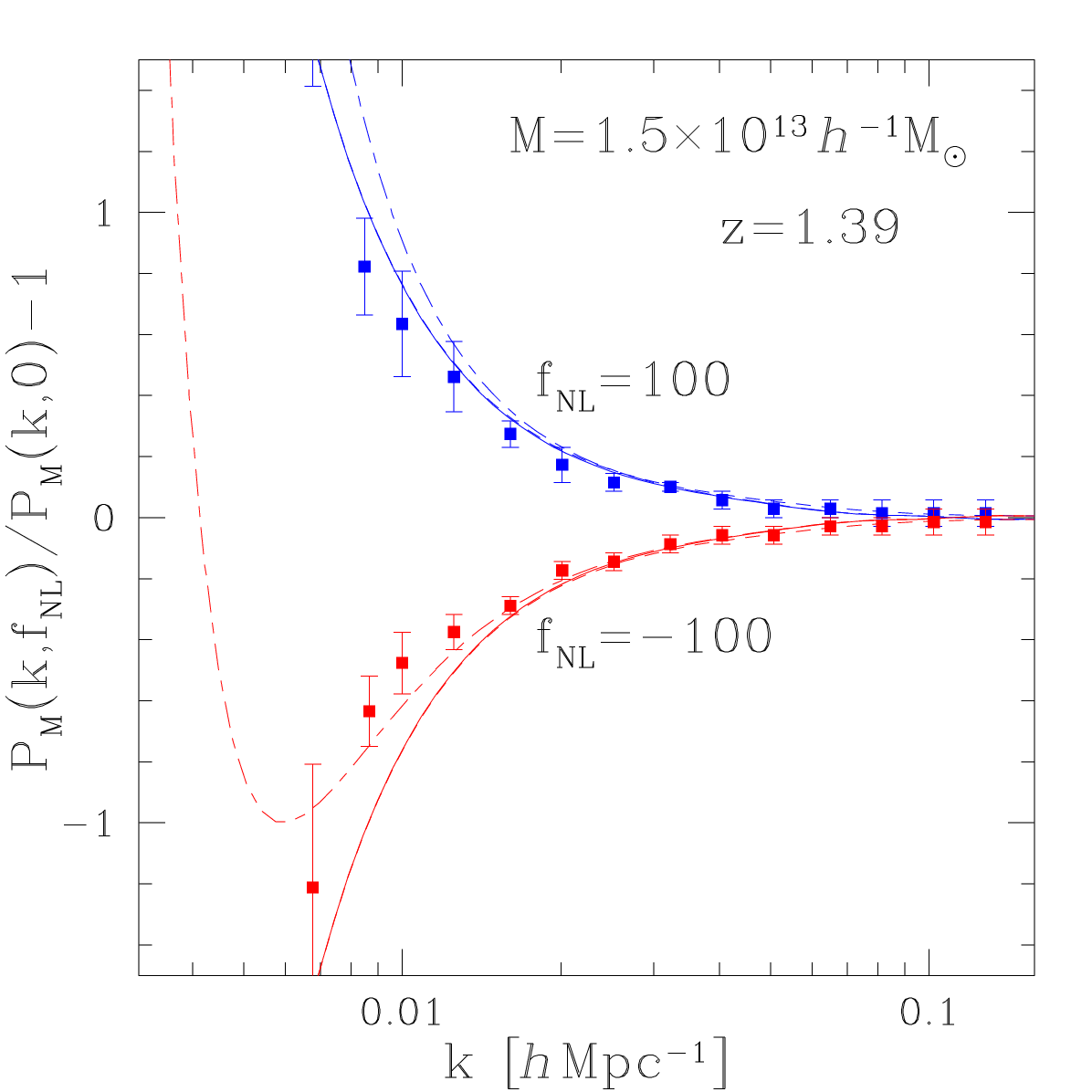}}
\epsfxsize=4.44 cm \epsfysize=4.5 cm {\epsfbox{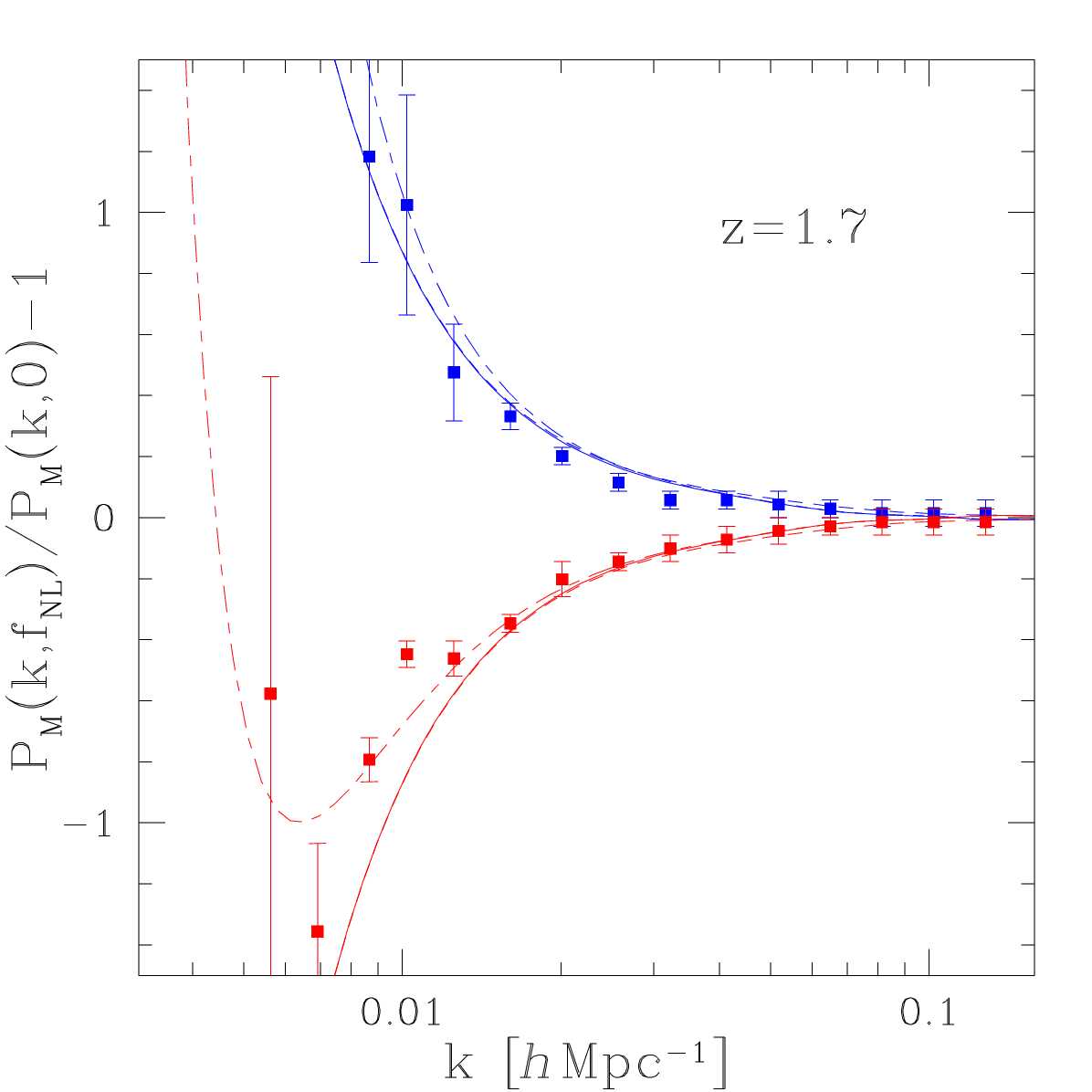}}\\
\epsfxsize=4.44 cm \epsfysize=4.5 cm {\epsfbox{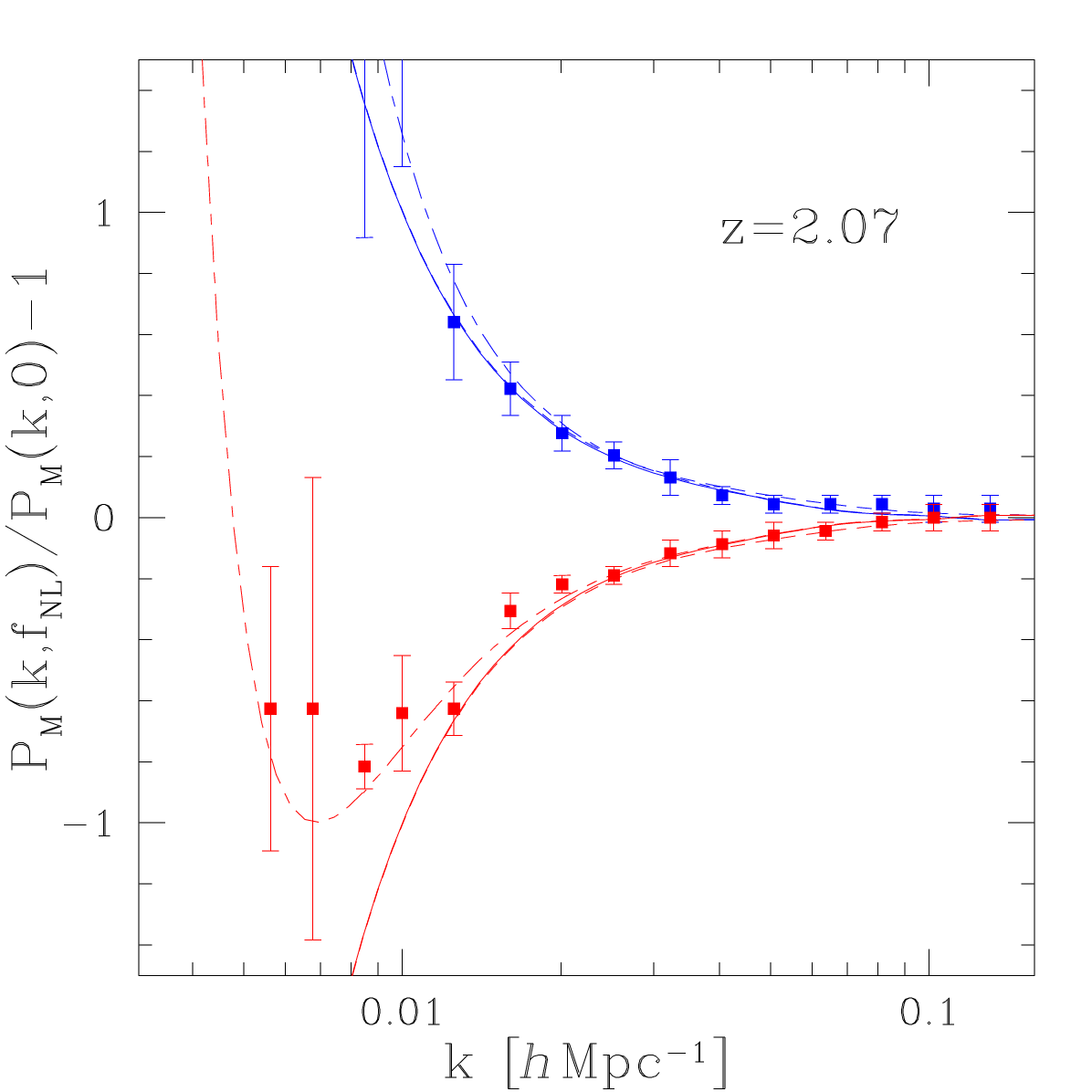}}
\end{center}
\caption{Same as Fig.~\ref{figbiask_M2d13_Desj}, but for
$M=1.5\times 10^{13}h^{-1} M_{\odot}$. The data points are the numerical simulations
of Desjacques et al. (2009), for $10^{13}<M<2\times 10^{13}h^{-1} M_{\odot}$.}
\label{figbiask_M1.5d13_Desj}
\end{figure}

If we take the limit of very rare events, which is $\sigma_q\rightarrow 0$
in Eq.(\ref{PkM}), we can only keep the last two terms (note that
$\tf \propto \sigma_q^2$, see Eq.(\ref{f112asymp}) below),
\beqa
\sigma_q \rightarrow 0 : \;\; P_M(k) & \sim & P_L(k) \tW(kq)^2 \left[ 
\frac{\delta_L^2}{\sigma_q^4} - 6 \frac{\delta_L^3 \, f_{1;11}}{\sigma_q^8}
\right] \nonumber \\
&& + 2 \frac{\delta_L^3}{\sigma_q^6} \left( \tf_{2;11}(k) + 2 \tf_{1;12}(k) 
\right) .
\label{PkM1}
\eeqa
At low $k$, with $\tW(kq)\simeq 1$, and at linear order over $\fNL$, the square
root of Eq.(\ref{bkdef}) gives with Eq.(\ref{PkM1}) the bias
\beqa
\sigma_q \rightarrow 0 , \;\; k\rightarrow 0 : \;\; b_M(k) & \sim & 
\frac{\delta_L}{\sigma_q^2} - 3 \frac{\delta_L^2}{\sigma_q^6} f_{1;11} 
\nonumber \\
&& + \frac{\delta_L^2}{\sigma_q^4} \frac{\tf_{2;11}(k) 
+ 2 \tf_{1;12}(k)}{P_L(k)} .
\label{bMk}
\eeqa
The first term, $\delta_L/\sigma_q^2$, is the result obtained by Kaiser (1984)
for rare massive halos (i.e. with a strong bias) for Gaussian initial conditions.
It is interesting to note that the second term gives a scale-independent
correction to the Gaussian bias. This term has already been noticed
in Slosar et al. (2008) and Afshordi \& Tolley (2008). As stressed in Desjacques
et al. (2009), taking this term into account is required to obtain a good match
to numerical simulations. We checked that this is indeed the case to match
the numerical results in
Figs.~\ref{figbiask_M2d13_Desj} and \ref{figbiask_M1.5d13_Desj} below
using Eq.(\ref{PkM}).
The last two terms depend on the wavenumber $k$.
There is an additional term that we neglect in this
paper, which arises from the dependence of the matter power spectrum on $\fNL$
(i.e. the denominator in Eq.(\ref{bkdef})). However, since this term is much smaller
than the other ones, we disregard it here (see Desjacques et al. 2009).
For the local non-Gaussianity (\ref{tfNLd_local}), we find from
Eqs.(\ref{tfabb})-(\ref{tfbab})
\beqa
k\rightarrow 0 : \;\; \tf_{2;11}(k) & \sim & \fNL \, \alpha(k) \int\dd\bk_1 \,
P_L(k_1)^2 \frac{\tW(k_1 q)^2}{\alpha(k_1)^2} , \nonumber \\
\tf_{1;12}(k) & \sim & \fNL \, \frac{\sigma_q^2 \, P_L(k)}{\alpha(k)} ,
\label{f112asymp}
\eeqa
which yields from Eq.(\ref{bMk})
\beq
\sigma_q \rightarrow 0 , \;\; k\rightarrow 0 : \;\; b_M(k) \sim 
\frac{\delta_L}{\sigma_q^2} - 3 \frac{\delta_L^2}{\sigma_q^6} f_{1;11} 
+ \fNL \frac{2 \delta_L^2}{\sigma_q^2 \alpha(k)} .
\label{bMk1}
\eeq
Thus, we recover the $k^{-2}$ dependence at low $k$ brought by the
local-type non-Gaussianity (\ref{fNLdef}), through the $1/\alpha(k)$ factor
in the last term. The second term in Eq.(\ref{bMk1}) is the constant shift
due to the non-Gaussianity noticed above.

In spite of the $k^{-2}$ dependence at low $k$ obtained in
Eq.(\ref{bMk1}) for the halo bias, the real-space halo two-point correlation is
well-defined and finite, as seen in section~\ref{saddle-point2}. 
Indeed, Eq.(\ref{bMk1}) only applies to a limited range, and one cannot
write the real-space two-point correlation as a Fourier transform of the form
$\xi_M (x) \sim \int \dd\bk \, e^{-\ii\bk.\bx} [1+\fNL/\alpha(k)]^2 P_L(k)$,
which would diverge at low $k$.
Thus, the advantage of the real-space approaches, such as the one described in this
paper, is that we obtain well-defined results in both real space and Fourier space,
and we do not need to regularize integrals by introducing a counterterm
associated with a survey-size window, as in Wands \& Slosar (2009).
This is reassuring, since one does not expect the halo correlation on a given scale
to depend on the size of the survey. Mathematically, the lack of divergence in
our approach comes from the fact that it is the halo power spectrum itself which
contains a term of the form $\Delta P_M(k) \sim \fNL P_L(k) /\alpha(k)$, see
Eqs.(\ref{PkM}) and (\ref{f112asymp}),
and it is only by expanding the square-root as in (\ref{Dbk}), 
$\Delta b = b [ \sqrt{1+  \Delta P_M/P_M}-1] \simeq b  \Delta P_M/(2P_M)$,
that $b$ can be written as in Eq.(\ref{bMk1}).
As we shall see below, in Figs.~\ref{figbiask_M2d13_Desj} and
\ref{figbiask_M1.5d13_Desj}, the expression (\ref{PkM}) is sufficient to explain the
behavior observed in numerical simulations, without introducing worrying
divergences.

We compare in Figs.~\ref{figbiask_M2d13_Desj} and \ref{figbiask_M1.5d13_Desj} our
results for the Fourier-space bias $b_M^2(k)$ with numerical simulations from
Desjacques et al. (2009).
Since the dependence on mass is rather weak, we show our results in
Fig.~\ref{figbiask_M2d13_Desj} for $M=2\times 10^{13}h^{-1} M_{\odot}$,
whereas the data points are for $M>2\times 10^{13}h^{-1} M_{\odot}$, and we show
our results in Fig.~\ref{figbiask_M1.5d13_Desj} for
$M=1.5\times 10^{13}h^{-1} M_{\odot}$, whereas the data points are for
$10^{13}<M<2\times 10^{13}h^{-1} M_{\odot}$.
The predictions (\ref{PkM}) and (\ref{PkMs}) are almost
indistinguishable in this regime, and they agree reasonably well with the simulations,
except at low $k$ for $\fNL=-100$ where they give a negative halo power spectrum.
Equation (\ref{bk1}), gives a much better
fit to simulations at low $k$ for negative $\fNL$, as could be expected from the
fact that it always gives a positive halo power spectrum.
However, a priori one should not give too much weight to this improved accuracy
in this regime. Indeed, as is clear from Eqs.(\ref{Dbk})-(\ref{bk1}), the solid and
dot-dashed curves in Figs.~\ref{figbiask_M2d13_Desj}-\ref{figbiask_M1.5d13_Desj}
only differ by terms of order $\fNL^2$ and beyond. Since all our results have been
derived at linear order over $\fNL$, one can expect that Eq.(\ref{bk1}) does not
include all terms of order $\fNL^2$. Then, although for practical purposes it is
better to use Eq.(\ref{bk1}) in this regime (i.e. negative $\fNL$ at low $k$), it is still
useful to also consider Eq.(\ref{PkM}), as the deviation between both predictions
should give an estimate of the theoretical uncertainty.

In contrast to some previous approaches
(e.g., Grossi et al. 2009; Desjacques et al. 2009), the good agreement with
numerical simulations shown in Figs.~\ref{figbiask_M2d13_Desj} and
\ref{figbiask_M1.5d13_Desj} is obtained from Eq.(\ref{PkM}) without any
fitting parameter (such as the rescaling parameter $q$ in Grossi et al. 2009
or the mass function parameters in Desjacques et al. 2009).
As for the halo mass function studied in sect.~\ref{Mass-function}, the role of this
parameter is partly played by the use of the exact linear threshold
$\delta_L=\cF^{-1}(200)$, which is predicted by the spherical dynamics of the
rare-event saddle points.
This makes formulae such as Eq.(\ref{PkM}) fully predictive for any values of
cosmological parameters.

\begin{figure}[htb]
\begin{center}
\epsfxsize=4.44 cm \epsfysize=4.5 cm {\epsfbox{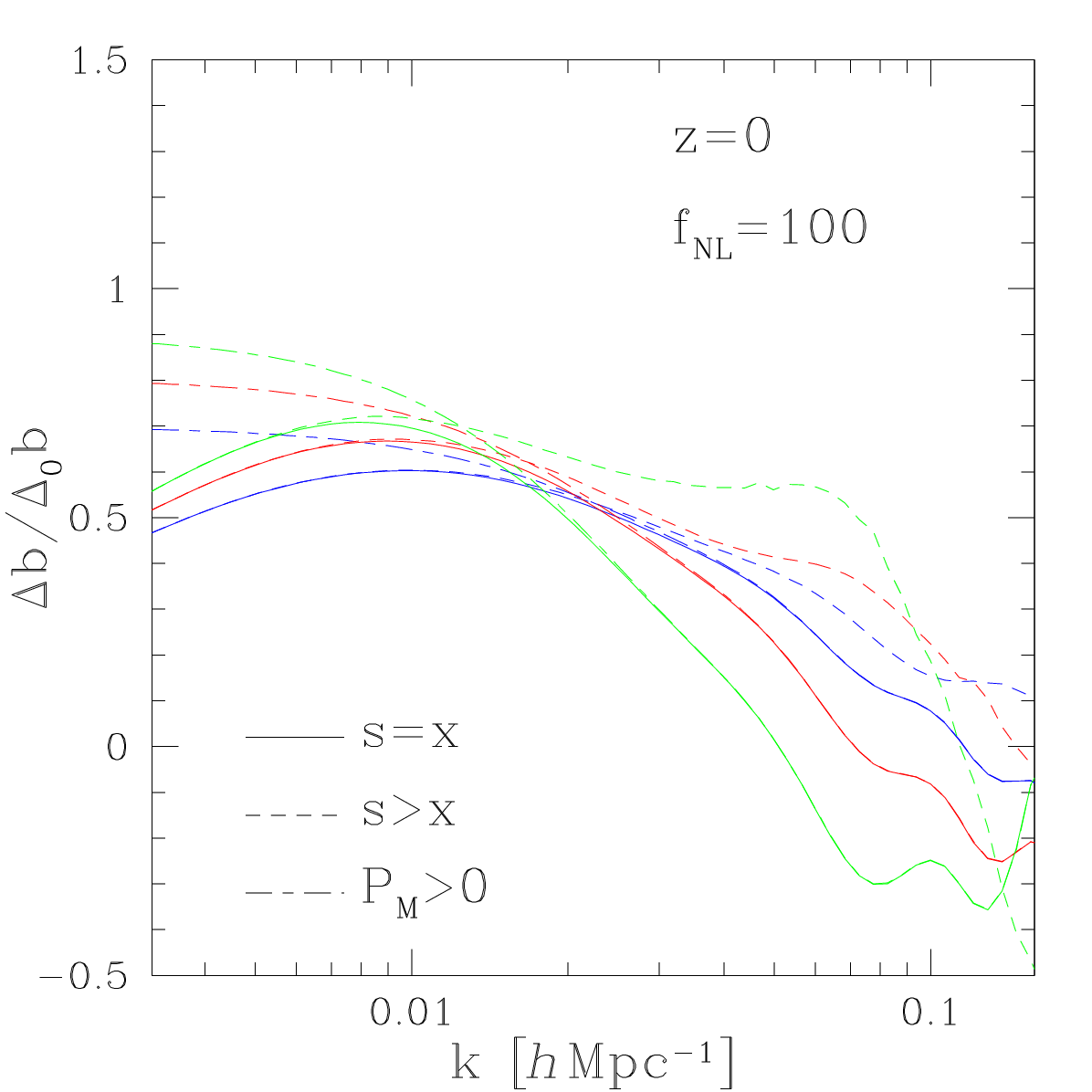}}
\epsfxsize=4.44 cm \epsfysize=4.5 cm {\epsfbox{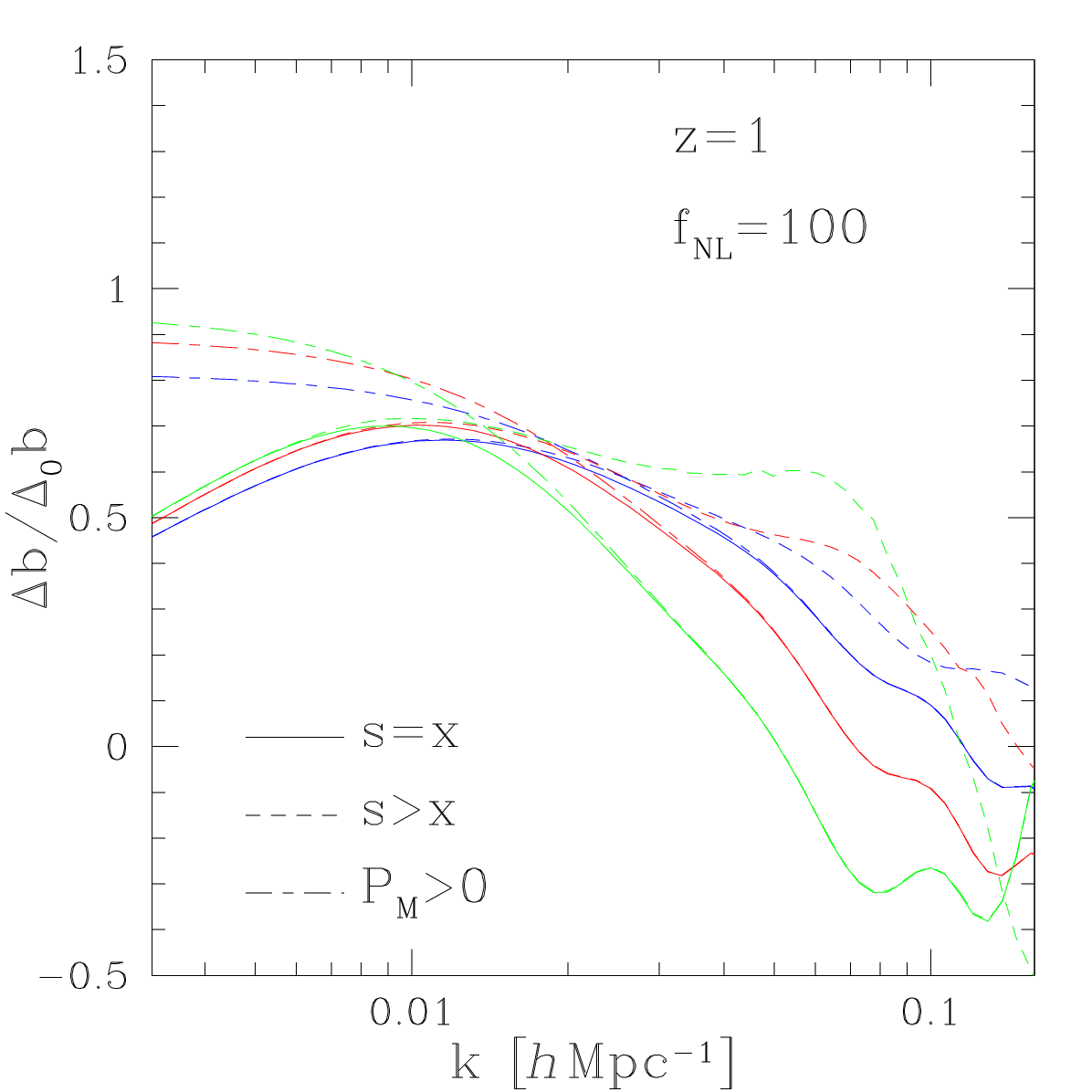}}
\end{center}
\caption{The Fourier-space ratio $\Delta b_M(k,\fNL)/\Delta_0 b_M(k,\fNL)$ of the
correction $\Delta b_M(k,\fNL)=b_M(k,\fNL)-b_M(k,0)$ to the Gaussian bias
$b_M(k,0)$, divided by the scaling factor $\Delta_0 b_M$ defined in
Eq.(\ref{Delta0bk}), as a function of wavenumber $k$.
We show the case $\fNL=100$ for several masses, $M=10^{13}$ (blue), $10^{14}$
(red), and $10^{15}h^{-1}M_{\odot}$ (green) at redshifts $z=0$ (left panel) and
$z=1$ (right panel). The curves labeled ``$s=x$'', ``$s>x$'', and ``$P_M>0$'',
correspond to Eqs.(\ref{PkM}), (\ref{PkMs}), and (\ref{bk1}), respectively.}
\label{figDeltabiask}
\end{figure}

As in Fig.~\ref{figDeltabiasr}, in order to see the scaling of the
correction $\Delta b_M(k,\fNL)=b_M(k,\fNL)-b_M(k,0)$ to
the Fourier-space Gaussian bias more clearly, we show the ratio
$\Delta b_M(k,\fNL)/\Delta_0 b_M(k,\fNL)$ in Fig.~\ref{figDeltabiask}, where we
now define
\beq
\Delta_0 b_M(k,\fNL) = \fNL \, b_M(k,0) \, \frac{2\delta_L}{\alpha(k)} .
\label{Delta0bk}
\eeq
The correction obtained from a simple peak-background split argument instead
gives $\Delta b_M(k,\fNL) = \fNL [b_M(k,0)-1] (2\delta_L/\alpha(k))$, as in
Eq.(\ref{Dbk_Dal}), see Dalal et al. (2008); Slosar et al. (2008).
However, in our formalism, both the leading term $\delta_L^2/\sigma_q^4$
(third term in Eq.(\ref{PkM})) and the subleading term $\delta_L/\sigma_q^2$
(first term in Eq.(\ref{PkM})) are modified by primordial non-Gaussianity
(the terms in the brackets that follow). The term $\delta_L/\sigma_q^2$ arises
from the Lagrangian to Eulerian space mapping (the prefactor
$(1+\delta_{LM})$ in Eq.(\ref{xiM1M2})), and it corresponds to the factor $1$
in the more usual Eq.(\ref{Dbk_Dal}), which is not modified in the simplest model.
However, in general we can expect non-Gaussianities to also affect this
Lagrangian to Eulerian space mapping, and our model gives an estimate of
this effect through Eq.(\ref{deltap1}). Therefore, we scale $\Delta_0 b_M(k,\fNL)$
with $b_M(k,0)$ rather than with $[b_M(k,0)-1]$ in Eq.(\ref{Delta0bk}).
We show the results obtained at redshifts $z=0$ and $z=1$ in Fig.~\ref{figDeltabiasr}
for several masses. The curves labeled ``$s=x$'', ``$s>x$'' and ``$P_M>0$''
correspond to Eqs.(\ref{PkM}), (\ref{PkMs}), and (\ref{bk1}) respectively.
We can see that the collapse of the curves obtained for different masses is
not exact, as could be expected since Eq.(\ref{PkM}) is more complex than
Eq.(\ref{Delta0bk}), but the scaling (\ref{Delta0bk}) still captures most of the
dependence on the halo mass $M$. In fact, most of the dispersion seen in
Fig.~\ref{figDeltabiasr} does not arise from the different masses but from the
various approximations (\ref{PkM}), (\ref{PkMs}), and (\ref{bk1}). This should
provide an estimate of the theoretical uncertainty. As in 
Figs.~\ref{figbiask_M2d13_Desj} and \ref{figbiask_M1.5d13_Desj}, Eq.(\ref{bk1}),
which involves higher-order terms over $\fNL$ to ensure that $P_M(k)$ is always
positive, rises above Eq.(\ref{PkM}) at low $k$ ($k<0.01h$ Mpc$^{-1}$) and is
indistinguishable at higher $k$, whereas Eq.(\ref{PkMs}) rises above (\ref{PkM})
at high $k$ ($k>0.01 h$ Mpc$^{-1}$), where the correction $\Delta b_M(k,\fNL)$ is
quite small, and is indistinguishable at lower $k$. Approximations (\ref{bk1})
and (\ref{PkMs}) are likely to be most accurate at low and high $k$, respectively,
see also Figs.~\ref{figbiask_M2d13_Desj} and \ref{figbiask_M1.5d13_Desj}.
In any case, we can see that the scaling with wavenumber,
$\Delta b_M(k,\fNL) \sim 1/\alpha(k)$, is slightly broken, although this still
provides a good approximation. As in the numerical simulations of
Desjacques et al. (2009), the ratio $\Delta b/\Delta_0 b$ is suppressed
at high $k$, which means that the correction $\Delta b_M(k,\fNL)$ decreases slightly
faster than $1/\alpha(k)$ over the range $0.01-0.1 h$ Mpc$^{-1}$.
In agreement with the real-space behavior seen in Fig.~\ref{figDeltabiasr},
it appears that the correction $\Delta b_M(k,\fNL)$ scales slightly more closely as
$k^{-2}$ than as $1/\alpha(k)$ over this range, although neither of these two behaviors
is exact.

\section{Conclusion}
\label{Conclusion}

We have shown in this article how to extend to non-Gaussian initial conditions
the computation of the mass function and of the bias of dark matter halos
presented in Valageas (2009b) for the Gaussian case.
This relies on a saddle-point approach that allows to derive the high-mass
asymptotic tails of the quantities of interest from the statistical weight of the
initial conditions, supplemented by additional nonlinear constraints.
Then, focusing on the case of ``$\fNL$-type'' primordial non-Gaussianity, where the
linear gravitational  potential can be written as the sum of linear and quadratic
terms over an auxiliary Gaussian field $\chi$, we explained how to obtain
the relevant saddle-points as a perturbative series over the nonlinear parameter
$\fNL$. This method is very general, and it applies to any case of small primordial
non-Gaussianity, where Bardeen's potential (or equivalently, below the Hubble
radius, the gravitational potential or the density field) 
can be written as a polynomial over a Gaussian field $\chi$ as
\beq
\Phi= \chi  + \sum_{i=2}^k \, f_i \left( \chi^i - \lag\chi^i\rag \right) , 
\label{chi_i}
\eeq
where the nonlinear parameters $f_i$ are small. Then, following the method
presented in sections~\ref{Mass-function-halos}, \ref{Bias-of-halos}, the one-cell
and two-cell saddle points (associated with the one-point and two-point 
density distributions, whence the halo mass function and bias) can be computed
as a perturbative series over the coefficients $f_i$, which need not be of the
same order. This includes the case of a cubic term $g_{\rm NL} \chi^3$ in particular.
Of course, it also extends to the cases where the coefficients $f_i$ are not
mere numbers but convolution kernels and to several Gaussian fields $\chi_j$
(which can have a nonzero cross-correlation).
In all such cases, the high-mass asymptotics are set by the statistical weight 
$e^{-(\chi_i.C_{ij}^{-1}.\chi_j)/2}$ of these Gaussian fields, taken at the saddle point
associated with the maximization of this weight under appropriate nonlinear
constraints that express the mapping from $\chi_j$ to the relevant quantity, such
as the nonlinear density within a spherical cell. 

Focusing on the case of local-type primordial non-Gaussianity, we described 
how to obtain, up to linear order over $\fNL$, the one-cell saddle point associated
with the probability distributions
$\cP_L(\delta_L)$ and $\cP(\delta)$ of the linear and nonlinear density contrasts
within spherical cells. This gives the quasi-linear limit of these distributions, as well
as the high-mass exponential falloff of the halo mass function. One advantage of
our method is that it allows us to explicitly check that realistic amounts of primordial
non-Gaussianity have no significant effect on the density profile of this saddle point.
This ensures that shell crossing appears for (almost) the same nonlinear density 
$\delta_+ \ga 200$ (see Valageas 2009b), so that the high-mass tail of the halo
mass function can be derived provided halos are defined by a nonlinear density
threshold that is below this upper bound (which is indeed the case).
Although this procedure only gives the high-mass tail, we proposed a
simple change of variable, applied to the mass function fitted to Gaussian numerical
simulations, that obeys this high-mass asymptotic while keeping the mass
function normalized to unity. If one uses the Press-Schechter mass function,
this gives back the result of Matarrese et al. (2000), but we argue that this procedure
is somewhat more natural if one wishes to recover a more accurate mass function
in the Gaussian case. We also checked that this agrees with results from
non-Gaussian numerical simulations.

Next, we applied this method to the two-point correlation of massive halos,
following the approach of Kaiser (1984). As in Valageas (2009b) we take the
displacement of halo pairs under their mutual gravitational attraction into account.
This gives the real-space halo correlation $\xi_M(x)$, whence the real-space bias
$b_M(x)$. Since this approach does not assume that the halo correlation is weak,
the nonlinear formula it yields can be used to check whether the ``linearized'' form
(where one only keeps the linear term over the matter correlation $\xi(x)$)
is a good approximation in the regime of interest. As expected, we find that
the correction $\Delta b_M(x,\fNL)$ to the Gaussian bias grows with $b_M(x,0)$
and with scale, roughly as $\fNL b_M(x,0) x^2$, up to $x \sim 100 h^{-1}$Mpc.
Beyond this scale, the baryon
acoustic oscillation and the fact that the halo and matter correlations do not change
sign at the same point lead to strong oscillations and divergent spikes for $b_M(x)$.
This means that, for $x > 100 h^{-1}$Mpc, the bias is no longer a useful quantity,
and one should directly work with the halo and matter correlations.
In agreement with Desjacques (2008), we find that the two-point correlation of
massive halos, which have a large bias, strongly amplifies the baryon acoustic
oscillation. In addition we also obtain the modifications associated with
primordial non-Gaussianity. The baryon oscillation remains strongly amplified,
with a small shift, but somewhat less so for positive $\fNL$.

Finally, we used the ``linearized'' form of the halo two-point correlation to
derive the halo power spectrum and the halo bias in Fourier space.
We also give a simple recipe that ensures that the halo power spectrum always
remains positive. (This only differs from the direct prediction by terms of order
$\fNL^2$ and higher.) We obtain good agreement with numerical simulations
without introducing any free parameter. Moreover, the two formulae described
above allow one to estimate the range over which linear approximations over
$\fNL$ are sufficient. Thus, we find that terms of order $\fNL^2$ start playing a role
at low $k$ ($k < 0.01 h$Mpc$^{-1}$) for large negative $\fNL$ ($\fNL < -100$),
where the direct formula would give a negative power spectrum.

We also pointed out that the $k^{-2}$ behavior observed at low $k$
for the halo bias does not imply any divergence for the real-space two-point
correlation. Indeed, this behavior is only obtained within a certain limit,
and it is the halo power spectrum itself (i.e. $\Delta b^2$ rather than
$\Delta b$) that shows this $k^{-2}$ factor. We showed that this is
sufficient to explain the behavior observed in numerical simulations.
Moreover, it avoids the need to introduce counterterms, that depend on the
size of the survey, so as to obtain finite real-space correlations. This is an advantage of
real-space approaches, such as the one presented in this paper, which are better
suited to describing the nonlinear effects associated with the bias of massive
halos.

These results, which do not involve free parameters except for the mass function
(if one requires its full shape, where one needs the fit to numerical simulations
for Gaussian initial conditions) should be useful for constraining primordial
non-Gaussianities from observations of large-scale structures.
Thus, neither the high-mass tail of the halo mass function nor the bias require
rescaling parameters (such as $\delta \rightarrow \delta_c\sqrt{q}$), because such a
correction to the linear threshold $\delta_c$ is achieved through the use of
the exact linear threshold $\delta_L=\cF^{-1}(200)$ predicted by the spherical
dynamics of rare-event saddle points. This makes this approach more predictive
than some of the previous works, since one does not need to run new simulations to
fit for such $q$-factors in order to investigate other cosmologies. 
In particular, as discussed above for Eq.(\ref{chi_i}), the method presented in this
article is quite general and can be applied to a large class of models.
Moreover, since it provides results in both real space and Fourier space
(i.e. the halo two-point correlation and power spectrum), it gives a complete and
consistent description of halo clustering. As for previous approaches, the most
reliable use of these models to constrain cosmology is to take advantage of the
specific shape of the dependence on mass (for the mass function) or scale
(for the bias) brought by primordial non-Gaussianity to constrain $\fNL$,
rather than the change in the amplitude at a given mass or scale.

\end{document}